\pdfoutput=1
\documentclass[useAMS,usenatbib]{mnras}
\usepackage{graphicx}
\usepackage{amsfonts}
\usepackage{amssymb}
\usepackage{amsmath}
\usepackage{color} 
\usepackage{url}
\usepackage{tikz}
\usepackage{gensymb}
\usepackage{newtxtext,newtxmath}

\def\idm#1{{\mbox{\scriptsize #1}}}
\def\tv#1{{\pmb #1}}

\title[Astrometric campaign of AM~Her with e-EVN]{Another look at AM~Herculis -- radio-astrometric campaign
with the~e-EVN at 6~cm}
\author[M. P. Gawro\'nski, K. Go\'zdziewski, K. Katarzy\'nski and G. Rycyk]
{M. P. Gawro\'nski\thanks{E-mail:motylek@astro.umk.pl}, K. Go\'zdziewski, K. Katarzy\'nski \& G. Rycyk \\
Centre for Astronomy, Faculty of Physics, Astronomy and Informatics, Nicolaus Copernicus University,\\
Grudziadzka 5, 87-100 Toru\'n, Poland\\}
\begin{document}

\date{Accepted 2017 December 01. Received 2017 August 31; in original form 2017 February 10.}

\pagerange{\pageref{firstpage}--\pageref{lastpage}} \pubyear{2018}

\maketitle

\label{firstpage}

\begin{abstract}
We conducted radio-interferometric observations of the well known binary cataclysmic  system AM~Herculis. 
This particular system is formed by the magnetic white dwarf (primary) and the red dwarf (secondary),  and is 
the prototype of so-called polars. Our observations were conducted with the European VLBI Network (EVN) in 
the e-EVN  mode at 5 GHz. We obtained six astrometric  measurements spanning one year, which make it possible 
to update the annual parallax for this system with the best precision to date ($\pi=11.29\pm0.08$\,mas), 
equivalent to the distance of $88.6\pm0.6$~pc.   The system was observed mostly in the quiescent phase 
(visual magnitude $m_v\sim15.3$),  when the radio emission was at the level  of about $300\,\rm{\mu}$Jy. 
Our analysis suggests that the radio flux of AM~Herculis is modulated with the orbital motion.  Such specific 
properties of the radiation can be explained using the emission mechanism similar to the scenario  proposed 
for  V471\,Tau and, in general, for RS\,CVn type stars. In this scenario the radio emission arises near 
the surface of the red dwarf,  where the global magnetic field strength may reach a few kG. We argue that 
the quiescent radio emission distinguishes AM\,Herculis  together with AR\,Ursae Majoris (the second known 
persistent radio polar) from other polars, as the systems with a magnetized  secondary star.
\end{abstract}

\begin{keywords}
star - astrometry: AM\,Her, cataclysmic variables - radio continuum: AM\, Her,
cataclysmic variables - radiation mechanisms: AM\,Her
\end{keywords}

\section{Introduction}

Cataclysmic variable stars are a broad class of binary systems that consist  a white dwarf (primary object) 
and mass transferring secondary star. The secondary component fills the Roche sphere and 
losses matter via L1 point. Among these objects we may distinguish a subclass of magnetic 
cataclysmic  binary systems -- called polars. 
In polars very strong magnetic field of the primary star (10\,--230\,MG) prevents
from the creation of accretion disks. The matter transferred from the secondary 
component must follow along magnetic field lines, and at the end falls into the
magnetic pole/poles of the primary star. Such mass transfer leads to the
formation of strong shocks nearby the surface of the white dwarf. As a result,  
the strong radiation is produced mainly by the bremsstrahlung and the 
electron cyclotron maser processes \citep{1982ApJ...259..844M, 1983ApJ...273..249D}. 
An irregular variability in polars' luminosity on timescales from days to months is one of  
their main characteristics. As polars have no accretion disc, the changes in luminosity reflect 
variations in the mass transfer rate.  The origin of the mass transfer rate instability is probably 
connected with  the local magnetic activity of the secondary component. When active region on 
the  secondary star is episodically drifting in the front of the inner Lagrangian point, 
the mass transfer is ceased due to the magnetic pressure of a large star-spot  
\citep{1994ApJ...427..956L,2005AJ....130..742K}. Alternatively, changes in the  mass 
transfer rate could reflect variations in the size of active chromosphere, when  
the secondary star is not fulfilling the Roche lobe \citep{2000ApJ...530..904H}.

\begin{table*}
\label{tab1}
\begin{center}
\begin{tabular}{l l l c c c c c c}
\hline
%\hline
\noalign{\smallskip}
Project  & \multicolumn{2}{c}{Date} &  Epoch     & \multicolumn{2}{c}{Conv beam}  
&  J1818+5017  & J1809+5007 & AM\,Her  \\
 code    & [day] & [UT]  & (JD-2450000) & [mas] & [deg]  & $S^{\rm core}_{5 {\rm GHz}}$ [mJy] 
&  $S^{\rm core}_{5 {\rm GHz}}$ [mJy] & $S_{5 {\rm GHz}}$ [$\mu$Jy] \\
\noalign{\smallskip}
\hline
 EG069A & 2012 Dec 5    &  07:06--09:47  &  6266.8521  &  10.0$\times$6.0 & -46  & 162$\pm$2 & 10$\pm$1   & 292$\pm$39 \\
 EG069B & 2013 Feb 6    &  04:37--07:26  &  6329.7517  &  8.4$\times$6.4  & -64  & 142$\pm$1 & 11$\pm$1   & 371$\pm$34 \\   
 EG069D & 2013 May 2/3  &  21:28--00:20  &  6415.4535  &  9.9$\times$6.2  & -40  & 174$\pm$1 & 11$\pm$1   & 244$\pm$29 \\
 EG069Ea & 2013 Sep 17   &  12:10--13:57  &  6553.0445  &  10.6$\times$5.4 & -51  & 166$\pm$2 & 9$\pm$1    & 178$\pm$32 \\
 EG069Eb & 2013 Sep 17   &  21:11--23:56  &  6553.4369  &  10.5$\times$5.5 &  56  & 167$\pm$1 & 13$\pm$1   & 347$\pm$31 \\   
 EG069F & 2013 Dec 3    &  16:38--19:05  &  6630.2444  &  9.9$\times$5.2  &  55  & 146$\pm$1 & 11$\pm$1   & 297$\pm$35 \\       
%\noalign{\smallskip}
%\hline  
%\noalign{\smallskip}
\hline
\end{tabular}
\end{center}  
\caption[]{The observational log of our astrometric campaign.}
\end{table*} 

\begin{table*}
\label{astr_data}
\begin{center}
\begin{tabular}{c c c c c c c c c }
%\noalign{\smallskip}
\hline
\noalign{\smallskip}
%\hline
  Epoch &  \multicolumn{4}{c}{AM\,Her}  &  \multicolumn{4}{c}{J1809+5007}  \\
(JD-2400000) &  $\alpha$\,(J2000)  & $\Delta$\,$\alpha$ [mas] & $\delta$\,(J2000)  
& $\Delta$\,$\delta$ [mas] & $\alpha$\,(J2000)  & $\Delta$\,$\alpha$ [mas] 
& $\delta$\,(J2000)& $\Delta$\,$\delta$ [mas]\\
\noalign{\smallskip}
\hline  
 47170.9996$^{*}$  &  18 16 13.30576  & 120    &  49 52 04.3330  & 120   & ---  & ---  & ---  & --- \\
 52929.5017$^{*}$  &  18 16 13.23569  & 120   &   49 52 04.9571  & 120   & ---  & ---  & ---  & --- \\
 
 56266.8521    &  18 16 13.192800 & 0.24  &  49 52 05.11172 & 0.23 & 18 09 15.069132  & 0.04 &  50 07 28.20070 &  0.04 \\
 56329.7517    &  18 16 13.193207 & 0.16  &  49 52 05.11928 & 0.15 & 18 09 15.069176  & 0.03 &  50 07 28.20091 &  0.02 \\     
 56415.4535    &  18 16 13.192216 & 0.20  &  49 52 05.14021 & 0.21 & 18 09 15.069147  & 0.03 &  50 07 28.20094 &  0.03 \\
 56553.0445    &  18 16 13.188295 & 0.31  &  49 52 05.14569 & 0.26 & 18 09 15.069123  & 0.07 &  50 07 28.20143 &  0.06 \\
 56553.4369    &  18 16 13.188338 & 0.18  &  49 52 05.14614 & 0.16 & 18 09 15.069160  & 0.02 &  50 07 28.20137 &  0.02 \\
 56630.2444    &  18 16 13.188045 & 0.24  &  49 52 05.14066 & 0.20 & 18 09 15.069169  & 0.02 &  50 07 28.20141 &  0.02 \\
\hline
\end{tabular}
\end{center}  
\caption[]{Astrometric position measurements of AM\,Her and J1809+5007. Their uncertainties are determined  from 
the {\sc AIPS} data fitting and do not include systematic effects. $*$ - positions based on archival  VLA data 
(experiments AC206 and AM783).}  
\label{tab2}
\end{table*} 

AM Herculis (hereafter AM\,Her) is one of the most intensively studied and 
intriguing magnetic cataclysmic binary system and the prototype of polars 
which are also named AM\,Herculis type stars. The irregular changes in AM\,Her 
brightness of $\Delta \mathrm{m}_v\simeq$\,2--3 mag (high and low states) are 
observed on timescales from weeks to months. \citet{1977ApJ...212L.125T} suggested
that AM\,Her contains a compact star with magnetic field of about 200~MG. 
After this AM\,Her became a frequent target in many observational campaigns,
conducted at different wavelengths \citep[e.g.][]{1978PASP...90...61S, 1978ApJ...220..261B, 1981ApJ...243..911F},  
including also the radio bands \citep{1982ApJ...259..844M, 1983ApJ...273..249D, 
1985ASSL..116..225B}. \citet{1991MNRAS.251P..37B} using infrared cyclotron features, 
estimated the white dwarf magnetic field strength to be $B\simeq14.5$\,MG, 
what was in a good agreement with previous results based on the Zeeman shifts of 
the photospheric absorption lines \citep[e.g.][]{1985MNRAS.212..353W}.
The inclination of the system orbit $i\simeq50^{\circ}$ \citep{1991MNRAS.251...28W,1996MNRAS.280..481D},
and the period $P\simeq3.094$\,hr \citep[e.g. ][]{1996MNRAS.280..481D, 2005AJ....130.2852K} were 
also derived. The white dwarf mass estimations are in the range between 0.35\,--\,1.0\,M$_{\odot}$
\citep[e.g. ][]{1993ASIC..403..411M,1998A&A...338..933G} with preferred value of 
$M_{wd}=0.6$\,--\,$0.7$\,M$_{\odot}$ \citep{1995ApJ...455..260W,1995A&A...303..127G}.
The secondary component is believed to be a M$4^+$--M$5^+$ type star 
\citep[e.g. ][]{1995A&A...303..127G,1995A&A...302...90S} with the mass $M_s=0.20$\,--\,$0.26$\,M$_{\odot}$
\citep{1995A&A...302...90S}.

\begin{figure}
\centering
\includegraphics[width=0.48\textwidth]{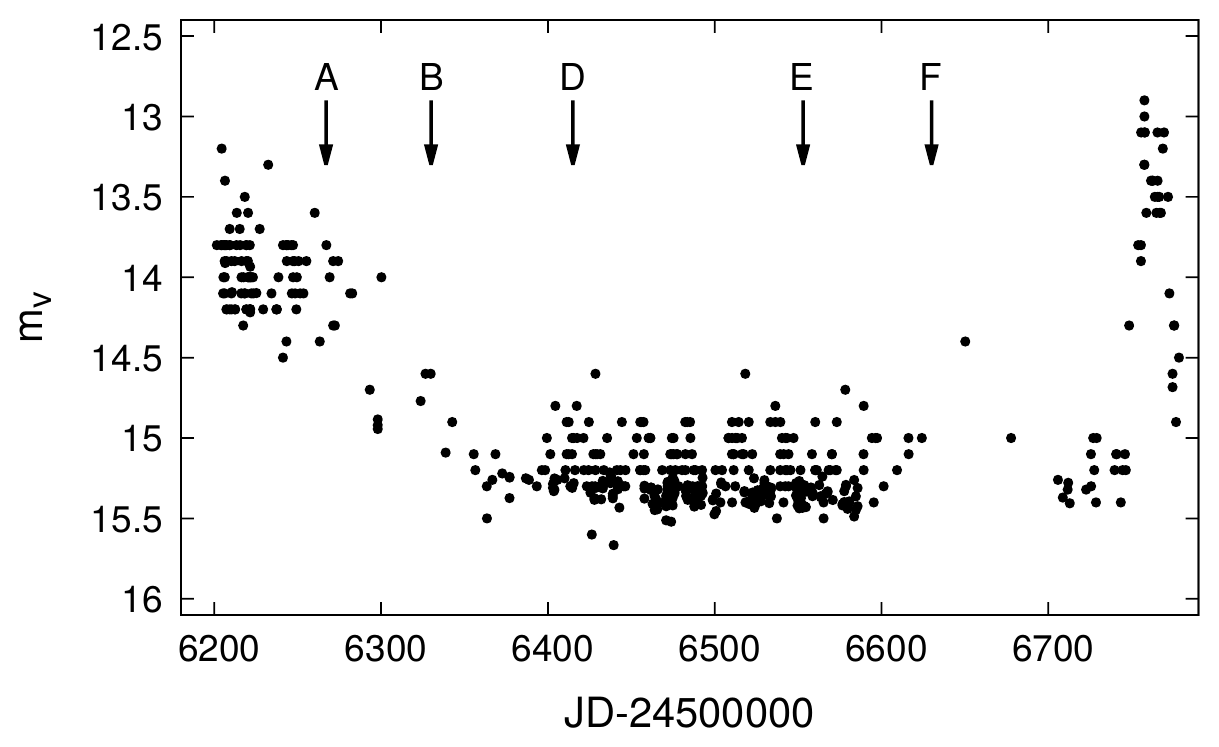} 
\caption[]{The optical light-curve of AM\,Her during our campaign (data from AAVSO).
The epochs of the e-EVN observations presented in this paper are indicated by arrows.}
\label{fig1}
\end{figure}
The first detection the AM\,Her radio emission was achieved using the Very Large Array (VLA) 
at 4.9~GHz \citep{1982ApJ...255L.107C}. The measured flux density of AM\,Her 
was 0.67 $\pm$ 0.05\,mJy, with no evidence of the circular polarization.  
\citet{1983ApJ...273..249D} confirmed this detection, obtaining the flux density 
of 0.55$\pm$0.05~mJy and also specifying the upper limits to the flux density at 1.4 and 
15~GHz (0.24\,mJy and 1.14~mJy, respectively). In addition, these authors also 
discovered the radio outburst at 4.9\,GHz with the maximum flux density of $9.7\pm2.3$ mJy, 
which  was $100\%$ RH circularly polarized. \citet{1983ApJ...273..249D} also attributed 
the quiescent emission to gyrosynchrotron process, caused by mildly relativistic 
electrons with energies $\sim$500\,keV, trapped in the magnetosphere of 
the white dwarf. The electron-cyclotron maser located nearby the red dwarf,
was proposed as a likely source of the radio outbursts. The same origin of radio 
flares was suggested by \citet{1982ApJ...259..844M}. 

\citet{1979ApJ...230..502Y} derived the first distance estimation to AM\,Her d$\simeq$75\,pc. 
Their finding was based on the analysis of various M-dwarf features in the optical spectrum.  
These authors additionally suggested, that secondary component in the system must be a M-dwarf 
with  the spectral type between M4 and M5. In the next paper \citet{1981ApJ...247..960Y} presented 
another AM\,Her distance estimation ($d=71\pm18$\,pc), using the near-infrared CCD spectra  
and TiO bands analysis, which also revealed the presence of M$4^+$ companion. However, there is 
an indication that the red dwarf is illuminated by the white dwarf and the spectral type of  
the secondary component may be modulated with the orbital phase \citep{1992MNRAS.257..476D}. 
\citet{1982AJ.....87..419D} determined the trigonometric parallax to 97 stellar systems, 
including AM\,Her ($d=108_{-28}^{+41}$\,pc). \citet{1995A&A...303..127G} used the 
so-called K-band surface-brightness method \citep{1981MNRAS.197...31B,1994ApJ...425..243R} 
and calculated the distance to AM Herculis as $91 _{-15}^{+18}$\,pc. The most recent 
distance estimation to the system of AM\,Her was made by \citet{2003AJ....126.3017T} 
with the use of the optical trigonometric parallax measurement with the 2.4\,m Hiltner 
Telescope ($d =79_{-6}^{+8}$\,pc).

In this paper we present a new astrometric campaign with the European VLBI\footnote{VLBI -- 
Very Large Baseline Interferometry} Network (EVN) at 6~cm wavelength, which was dedicated 
to the precise estimation of the AM\,Her annual parallax. This new value may be crucial for  
further modelling of physical processes in this system. The paper  is structured as follows. 
In \S~2 we describe observations and the data reduction. In \S~3 we present  a new astrometric 
model of AM\,Her. In \S~4 we discuss observed AM\,Her radio properties and the orbital phase 
dependence of the radio emission. Finally, in \S~5 we summarize our  conclusions. 

\begin{figure*}
\centering
\includegraphics[width=7.5cm]{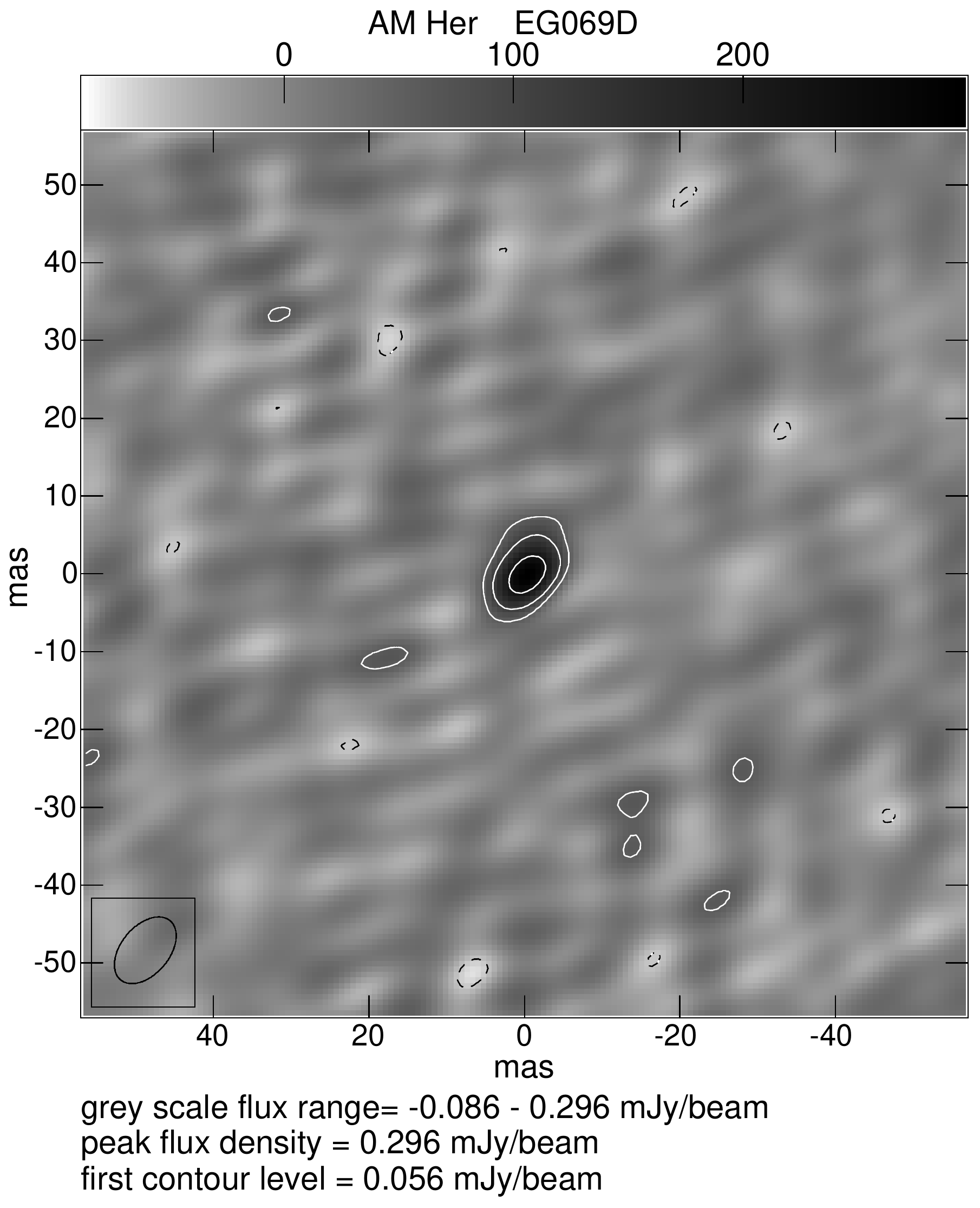} 
\includegraphics[width=7.5cm]{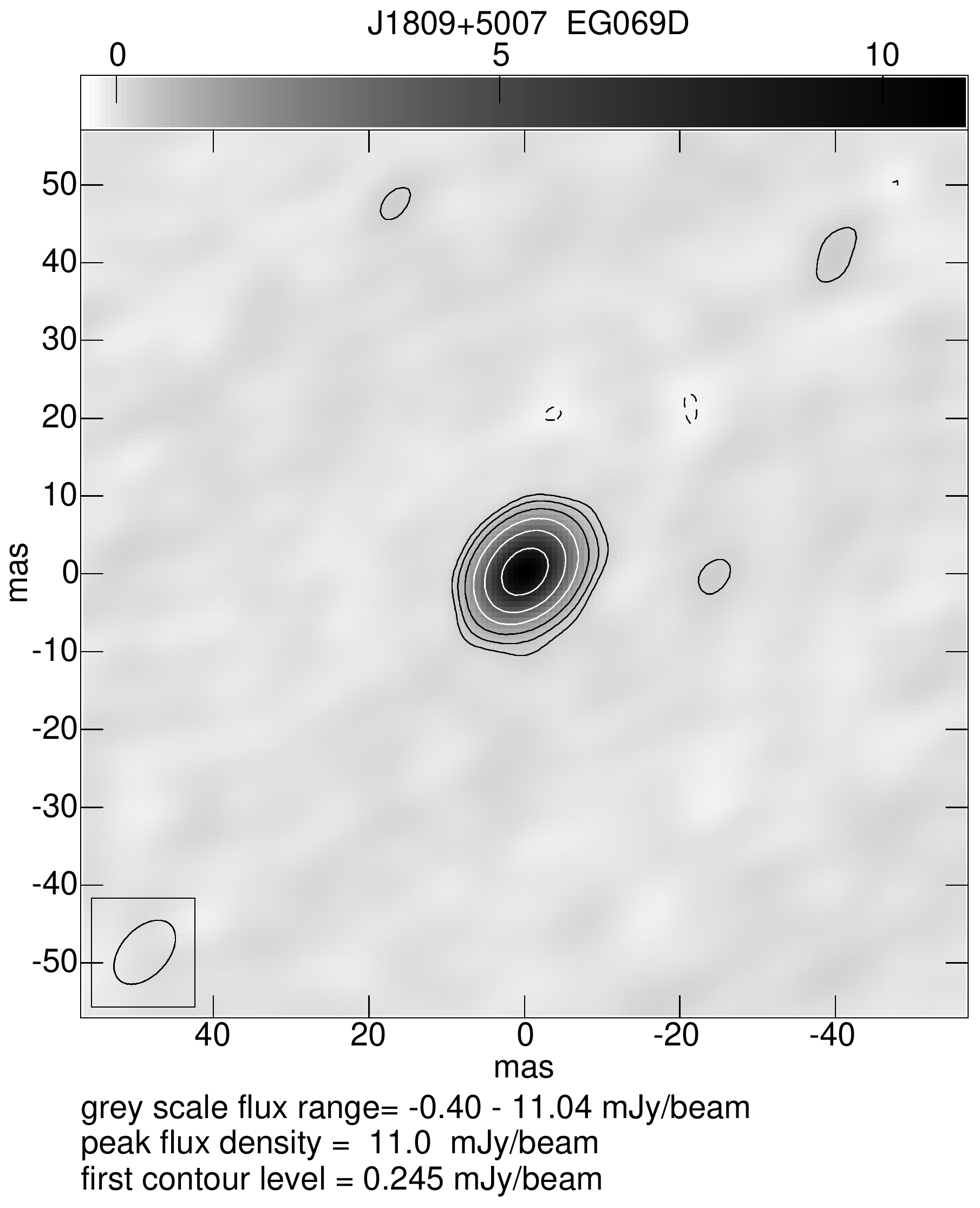}
\includegraphics[width=7.5cm]{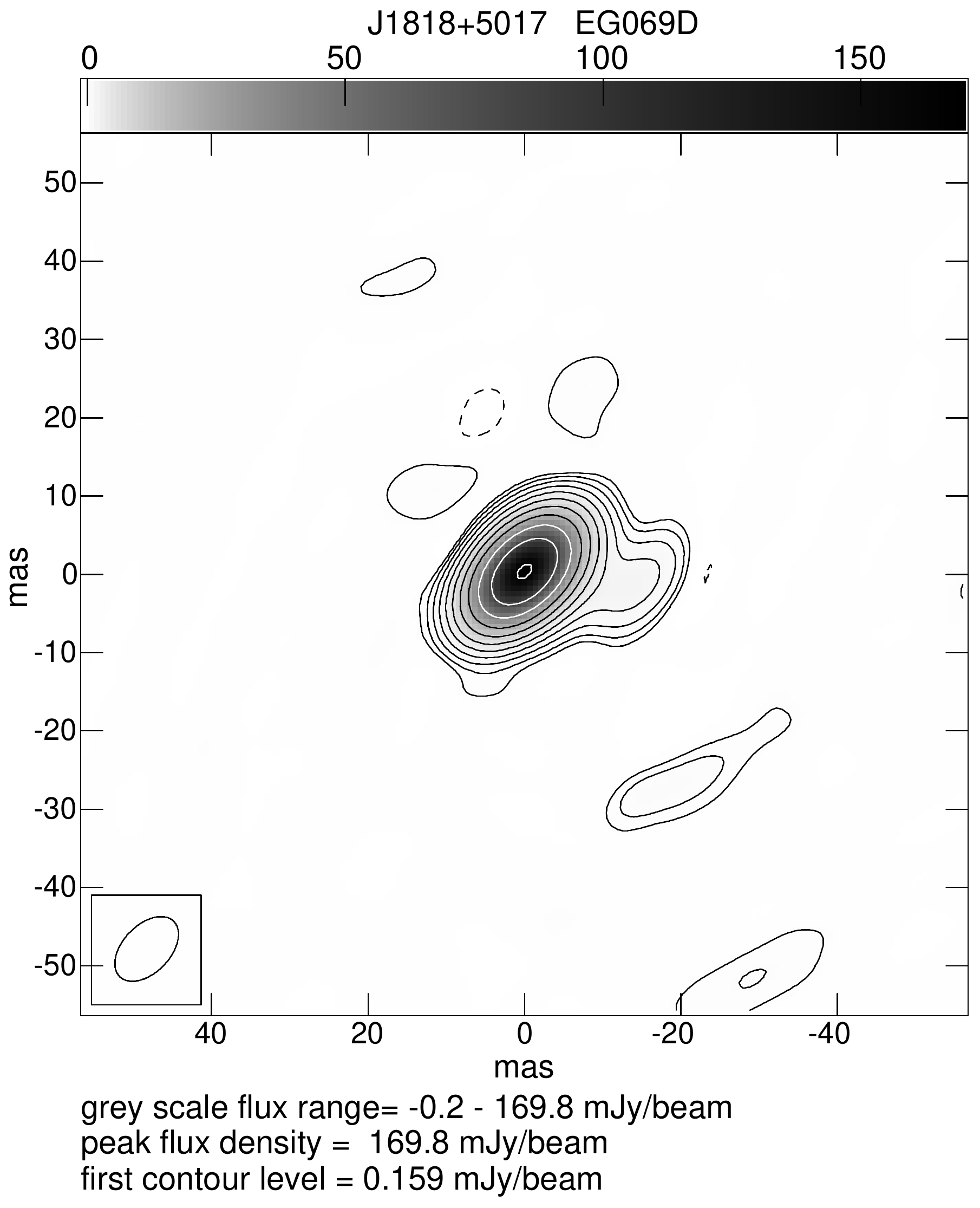} 
\includegraphics[width=7.5cm,trim=0 -2.7cm 0 0]{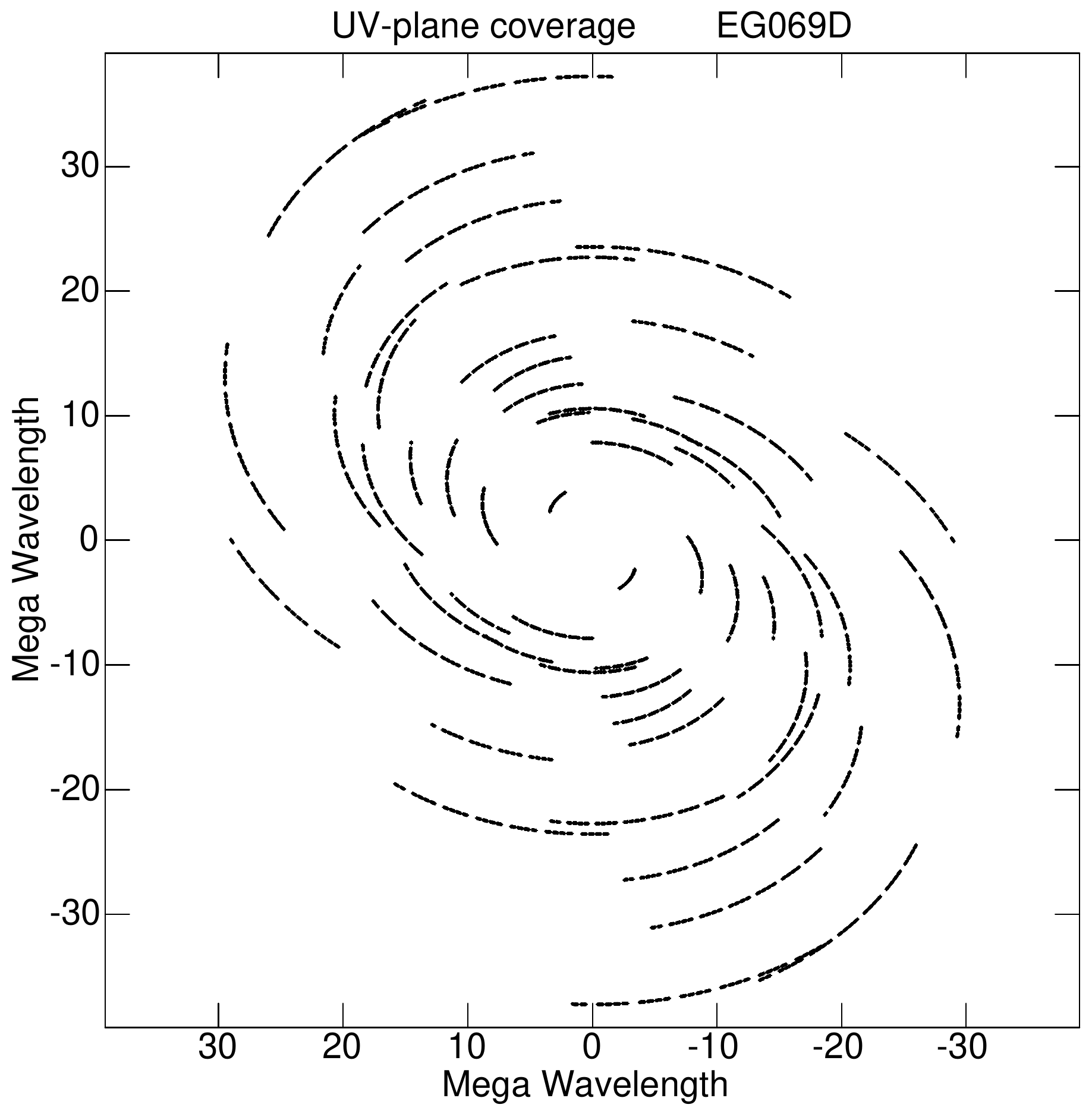}
\caption{An example of the radio maps obtained from our campaign (\emph{top left}: AM\,Her, 
\emph{top right}: J1809+5007, \emph{bottom left}: J1818+5017). The data were collected during 
the third epoch of the observations (EG069D). The successive  contours show an increase of 
the flux density by a factor of 2, where the first contour  corresponds to the detection limit 
of $\simeq\,3\sigma$. The insets show the size of the restoring  beam. The bottom right plot 
shows an example of typical AM\,Her observation \emph{uv}-plane  coverage, here also for EG069D 
part of the campaign.
}
\label{fig2}
\end{figure*}

\section{Observations and data reduction}

The interferometric observations of AM\,Her at the 5~GHz band were carried out in 
5 epochs, spread over 12 months from 2012 December 5 to 2013 December 3, using the EVN 
in the e-VLBI mode of observations, with use of the phase-referencing technique. The stations 
from Effelsberg, Jodrell Bank (MkII), Medicina, Noto, Onsala, Toru\'n, Yebes and Westerbork 
(phased array) participated in our observations (proposal code EG069).  The data were recorded 
at the rate 1~Gb/s providing a total bandwidth of 128\,MHz, divided into 8 base-band channels with 
a bandwidth of 16\,MHz each. The fourth epoch  was separated into two parts due to time allocation 
and both segments were treated during the reduction of data as separated epochs. The observation 
details of the all epochs are summarized in Tab.\ref{tab1}. 

Each observational epoch spanned $\sim$3 hours covering scans on: AM\,Her, a bright bandpass 
calibrator, the phase-reference source (J1818+5017) and the secondary calibrator (J1809+5007). 
During the astrometric calculations we used J1818+5017 position taken from 
Radio Fundamental Catalog\footnote{\url{astrogeo.org/rfc/}}, version \emph{rfc 2016d}
(RA=$18^{h}18^{\idm{m}}30^{\idm{s}}.519224$, DEC=$50^{\circ}17'19''.74353$, J2000.0).
It means that we corrected the original position measurements based on radio maps by 
-0.13\,mas in RA and -0.14\,mas in DEC, respectively. These shifts compensate 
the difference between J1818+5017 positions in catalogs \emph{rfc 2016d} and \emph{rfc 2012b} 
(\emph{rfc 2012b} was used during our observations and the correlation process).

J1809+5007 is a compact radio source selected from the Cosmic Lens All-Sky 
Survey\footnote{\url{www.jb.man.ac.uk/research/gravlens/class/class.html}}  \citep[e.g.][]{2003MNRAS.341....1M}
placed in  the proximity of AM\,Her (GB6J180913+500748, F$_{\mathrm{8.4\,GHz}}=25.2$\,mJy). 
J1809+5007 was observed to examine the phase-referencing success. The three observed sources 
are separated on the sky plane as follows: AM\,Her \& J1818+5017 
by 0\degree.6, AM\,Her \& J1809+5007 by 1\degree.2 and J1818+5017 \& J1809+5007 by 1\degree.5\,.
The observations were made in 5 minute long cycles, 3.5~minutes for AM\,Her or J1809+5007, and 1.5~minute 
for the phase calibrator. The main loop of the observations contains five such cycles. The first cycle 
dedicated to J1809+5007 was followed by four cycles, which included integrations on AM\,Her. It should 
be mentioned that the first two epochs of the  observations overlap with a decrease of the optical luminosity 
(blocks A \& B). The rest of the measurements were recorded at the optical low-state of AM\,Her  
(blocks D, E \& F). We show the AM\,Her optical light-curve during the campaign and the moments of 
our observations in Fig.\ref{fig1}. The visual optical observations are taken from American Association 
of Variable Star Observers\footnote{\emph{AAVSO}, \url{www.aavso.org}} database. 

The whole data reduction process was carried  out using standard NRAO package 
{\sc aips}\footnote{\url{www.aips.nrao.edu/index.shtml}} procedures \citep[e.g.][]{2003ASSL..285..109G}. 
The maps of the phase calibrator J1818+5017 were created with the self-calibration in the phase and the amplitude, 
and used as a model for final fringe-fitting.  We applied the {\sc imagr} task to produce the final total intensity 
images of all observed sources. During the mapping process the natural weighting was used. AM\,Her appears 
point-like at the radio maps, but for J1818+5017 we have detected a weak jet, pointed into east direction 
on all epochs. J1809+5007 also seems to be resolved at our maps with a hint of the jet suggestively directed 
into north-east. 

A sample of radio maps obtained during our observations is showed on Fig.\ref{fig2}. The radio fluxes and 
astrometric positions of all observed targets were then measured by fitting Gaussian models, using the 
{\sc aips} task {\sc jmfit}. In the case of J1809+5007 and J1818+5017,  we estimated only fluxes of 
the core, since the detailed modelling of these two sources is  beyond the scope of this work. Moreover, 
these core fluxes completely dominate over the  resolved structures. The flux variability was tracked 
using the task {\sc dftpl} with an averaging  interval equal to the length of scans. Before application 
of the {\sc dftpl} task, we searched the area  within the radius of 3\arcsec~around target's position 
for background sources and none was  found. In a case of the background object detection, its model 
should be removed from  \emph{uv}-data, before radio flux estimation with {\sc dftpl}. 

In order to expand the time span of the observations and to improve the proper motion estimation, we added  
two archival VLA observations at 8.4\,GHz, made  in BnA configuration (1988 Jan 10 and 2003 Oct 17,  
observational codes AC206 and AM783). The VLA observations were reduced with the {\sc aips}  package. 
Sources J1808+4542 and J1800+7828 were used as phase calibrators during AM783 and AC206,  respectively. 
All positions collected in this paper are presented in Tab.\ref{tab2}.  The used VLA observations where 
short scans and the distant phase calibrators were used.  Under the typical conditions of VLA observations, 
the astrometry accuracy of  $\sim$10\,\% of the  restoring beam could be achieved ($\sim$0.$''$1 for 
AC206 and AM783). In the archival observations,  the conditions are worse than during the standard 
observation, hence we made a crude estimation of the systematic error 0.$''$12 for VLA astrometry. 
It should be noted that VLA calibrators positions used during observations are taken 
from the different catalogs. This results in additional systematic effects in astrometry, which in general
should be taken into account during calculations as calibrators position are relevant to the different
global astrometric solutions. However, this effect is at the level of $\sim1$\,mas (the typical discrepancy
between a given source position in different catalogs are sub-mas), and hence negligible in comparison 
to assumed by us systematic errors for VLA position measurements.

\section{Astrometric model and an estimate of the absolute parallax}
Given the e-EVN measurements in the geocentric frame, we determine the parallax and the components of  the 
proper motion through a canonical 5-element model for the ICRS astrometric place of an isolated target:
\begin{equation}
\vec{r}(t_i) = \vec{r}(t_0) + \vec{m}(t_i-t_0) - \pi\vec{E_{\rm B}}(t_i),
\label{eq:m1}
\end{equation}
where $\vec{r}(t_i)$ is the geocentric position of the target at epoch $t_i$, relative to the  position vector $\vec{r}(t_0)$ in a reference epoch $t_0$, $\vec{m}$ is the space motion vector,  $\vec{E_{\idm B}}(t_i)$ is the barycentric position of the Earth at the observational epoch and  $\pi$ is the parallax of the radiation source. We note that due to a proximity of  the phase-calibrators and the target, we skip local, differential perturbations, for instance due  to the light deflection. We compute components of the vector mean motion  $\vec{m} \equiv [ m_{\idm{x}}, m_{\idm{y}}, m_{\idm{z}}]$ by fixing the radial velocity  $V_{\idm{R}}=-12\,\mbox{km\,s}^{-1}$ \citep{1979ApJ...230..502Y}.

The vector model in Eq.~\ref{eq:m1} may be parametrized through the target's ICRF coordinates  ($\alpha_{\rm 0},\delta_{\rm 0}$) at the initial epoch $t_{\rm 0}$,  components $(\mu_\alpha,\mu_\delta)$  of the proper motion at  the epoch $t_{\rm 0}$, and  the parallax factors $(\pi_\alpha,\pi_\delta)$  projected onto the ICRF coordinates axes. To avoid correlations between the zero-epoch position and  the proper motion components, we calculate the reference epoch $t_{\rm 0}$
\begin{equation}
t_{\rm 0} = \frac{\sum_i^{M} t_i w_i}{\sum_i^{M} w_i} \equiv \mbox{JD~2456457.5}, \quad w_i=\frac{1}{\sigma_i},
\label{eq:t0}
\end{equation}
which is the weighted mean of the observation epochs $t_i$, $i=1\ldots,M$, and $\sigma_i$ are formal  uncertainties  of the astrometric positions of the target derived from the radio maps. The barycentric  position of the Earth was determined in accord with the Solar system ephemeris JPL DE405 \citep{2014IPNPR.196C...1F}.

We optimized the astrometric model with the same approach as presented in \citet{2017MNRAS.466.4211G}.  Uncertainties of the model parameters $\tv{\xi} = (\alpha_0,\delta_0,\mu_\alpha,\mu_\delta,\pi)$ depend  on complex way on the residual phase in phase-referencing, sub-mas changes of the phase calibrator radio structures and the atmospheric zenith delay residuals.  To account for such factors, the formal 
uncertainties are rescaled in quadrature, $\sigma^2_i \rightarrow \sigma^2_i+\sigma_f^2$, where $\sigma_f$ is the so called error floor added as an additional free parameter to be optimized. To do so, we define the maximum likelihood function ${\cal L}$.  For normally distributed uncertainties $\sigma_i$, we account for the error floor $\sigma_f$, defining ${\cal L}$ as follows
\begin{equation}
 \log {\cal L} =  
-\frac{1}{2} \sum_{i,t} \frac{\mbox{(O-C)}_i^2}{{\sigma_i}^2+\sigma_f^2}
- \frac{1}{2}\sum_{i} \log ({{\sigma_i}^2+\sigma_f^2})  - M \log{2{\rm \pi}},
\label{eq:Lfun}
\end{equation}
where $(\mbox{O-C})_{j,t}$ is the (O-C) deviation of the observed $\alpha(t_i)$ or $\delta(t_i)$ at  epoch $t_i$ from its astrometric ephemeris  in Eq.~\ref{eq:m1},   for $i=1\ldots,M$ where $M$ is  the total number of $\alpha(t_i)$ and $\delta(t_i)$ measurements. It makes it possible to determine the error floor $\sigma_f$ in a self-consisted manner. 

We analyse the $\log{\cal L}$ function in terms of the Bayesian inference. We sample the posterior  probability distribution ${\cal P}(\tv{\xi}|{\cal D})$ of astrometric  model parameters $\tv{\xi}$  in Eq.~\ref{eq:m1}. Given the data set ${\cal D}$ of astrometric observations data-set (understood  as $\alpha_i$ and $\delta_i$ components): 
$
  {\cal P}(\tv{\xi}|{\cal D}) 
  \propto {\cal P}(\tv{\xi}) \, {\cal P}({\cal D}|\tv{\xi}),
$
where ${\cal P}(\tv{\xi})$ is the prior, and the sampling data distribution  ${\cal P}({\cal D}|\tv{\xi}) \equiv \log{\cal L}(\tv{\xi},{\cal D})$. For all parameters, we  define noninformative priors by constraining the model parameters, i.e.,  $\alpha_{\rm 0}>0$~hr, $\delta_{\rm 0}>0$~deg, $\mu^*_{\alpha}, \mu_\delta \in [-1000,1000]$~mas yr$^{-1}$, $\pi>0$~mas and $\sigma_f>0$~mas.

We used the {\sc emcee} package developed by \citet{Foreman2014} to perform the posterior sampling with the Markov Chain Monte Carlo (MCMC) technique. In all experiments, we increased the MCMC chain lengths to up to 72,000 samples and 2240 random ``walkers'' selected initially in a small radius hyperball around a preliminary astrometric solution found with the simplex algorithm. The MCMC acceptance ratio was near 0.5 in all cases.

We performed a few fitting experiments for three sets of measurements (see Table~\ref{tab2}). The first set comprises of all six EVN detections. The second set contains all EVN and VLA epochs. The third set comprises of a minimal number of four EVN epochs that make it possible to determine the absolute parallax. We optimized the 5-element model with and without the error floor. For the EVN data, we computed the astrometric parameters at the GAIA DR1 epoch JD~2457023.5, to have a direct link to the forthcoming GAIA catalogue \citep{GaiaDR1}.

\begin{figure*}
    \centering
    \includegraphics[width=0.86\textwidth]{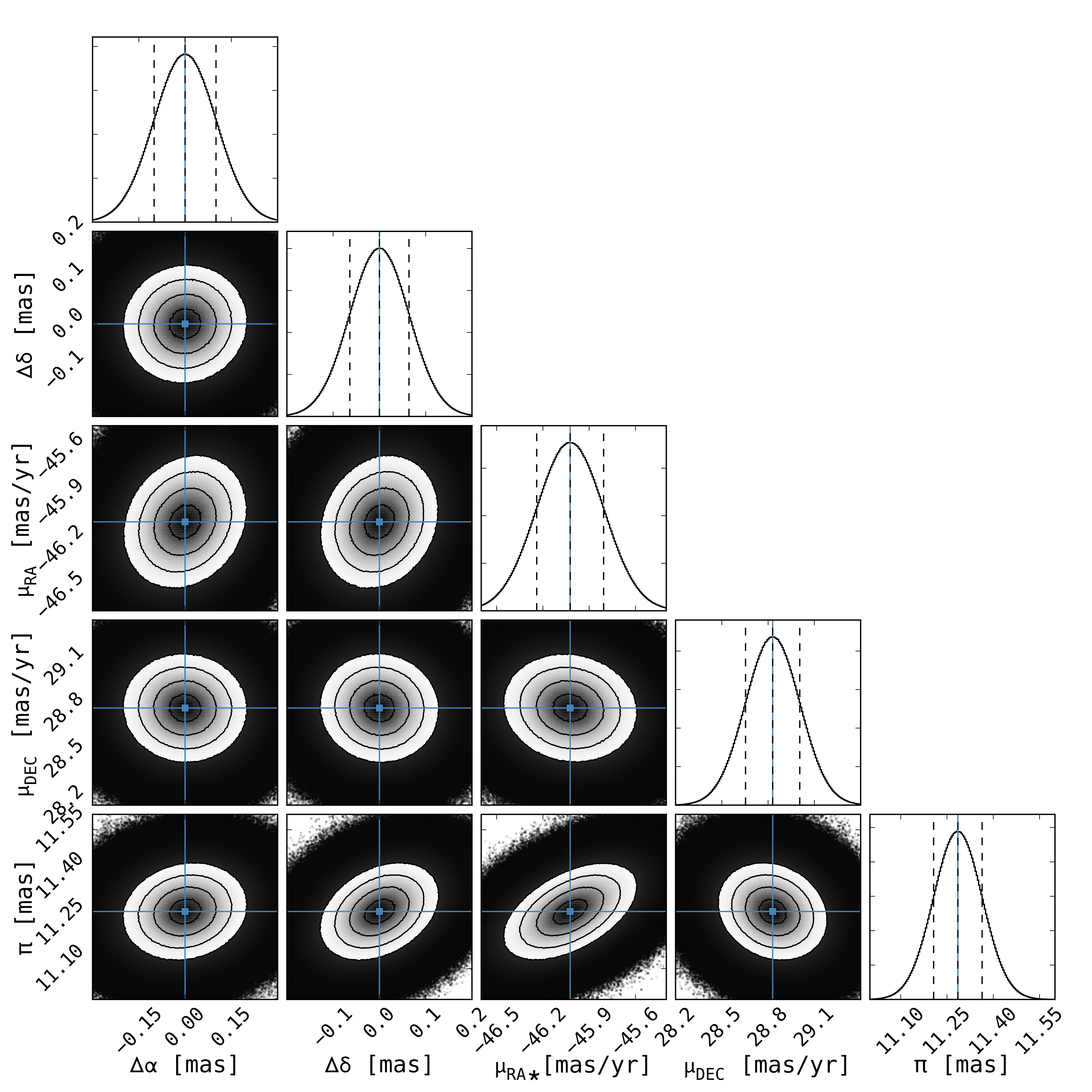}    
    \caption{
One- and two-dimensional projections of the posterior probability  distributions  of the astrometric best-fitting parameters for all EVN detections (Tab.~\ref{tab2}) expressed through the median values and marked with  the crossing blue/gray lines. The $\Delta\alpha_{\rm 0}$ and $\Delta\delta_{\rm 0}$   represents  offsets relative to the position at the reference epoch $t_{\rm 0}=\mbox{JD~2456457.5}$. Contours indicate 16th, 50th, and 84th  percentiles of the samples in the posterior distributions also marked between vertical, dotted lines in 1-dim histograms.  A single quasi-Gaussian  peak of the posterior appears clearly for all parameters. The model parameters do not exhibit strong pairwise correlations, though weak, near-linear correlations between $\pi$ and $\mu_{RA*} \equiv \mu^*_{\alpha}$, as well as between $\pi$ and $\Delta{\delta}$ are apparent.
}
\label{fig3}
\end{figure*}

\begin{figure*}
    \centering
    \includegraphics[width=0.86\textwidth]{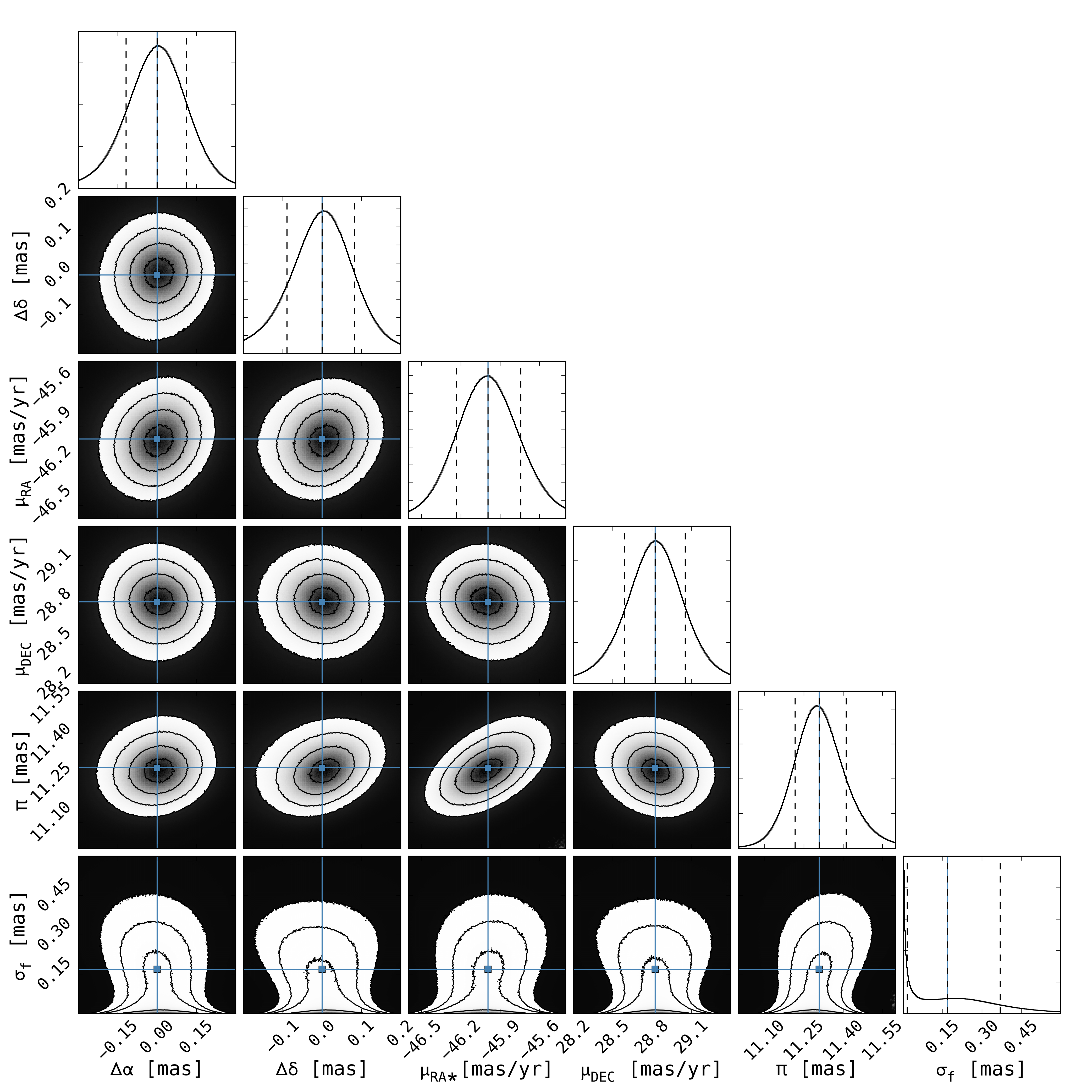}    
    \caption{
The same, as in Fig.~\ref{fig3}, but for six parameters, this time including the error floor parameter ($\sigma_f$). Note that parameter ranges in both plots are the same.
}
\label{fig3a}
\end{figure*}

The derived astrometric parameters for different data sets and epochs are displayed in Table~\ref{tab3}. Since the MCMC posteriors look  similar for all datasets, we illustrate only the  astrometric model for six EVN epochs (see Tab.~\ref{tab2} and the left column in Table~\ref{tab3}). The results are illustrated in  Fig.~\ref{fig3} and Fig.~\ref{fig3a}  as one-- and  two-- dimensional projections of the posterior probability distributions for astrometric model with and without accounting for the error floor correction, respectively. It may be compared with the posterior derived for a model with the error floor included (Fig.~\ref{fig3a}). The error floor parameter is in fact redundant for the AM Her e-EVN data, since its posterior probability has a maximum at $\sigma_f= 0$~mas, it is small and almost ``flat'' elsewhere with a median around 0.15~mas. Indeed, the 5-parameter astrometric model yields $\chi^2\sim 1$, and including the additional parameter does not improve the astrometric fit. 

\begin{figure}
\centerline{
\vbox{
    \includegraphics[width=0.48\textwidth]{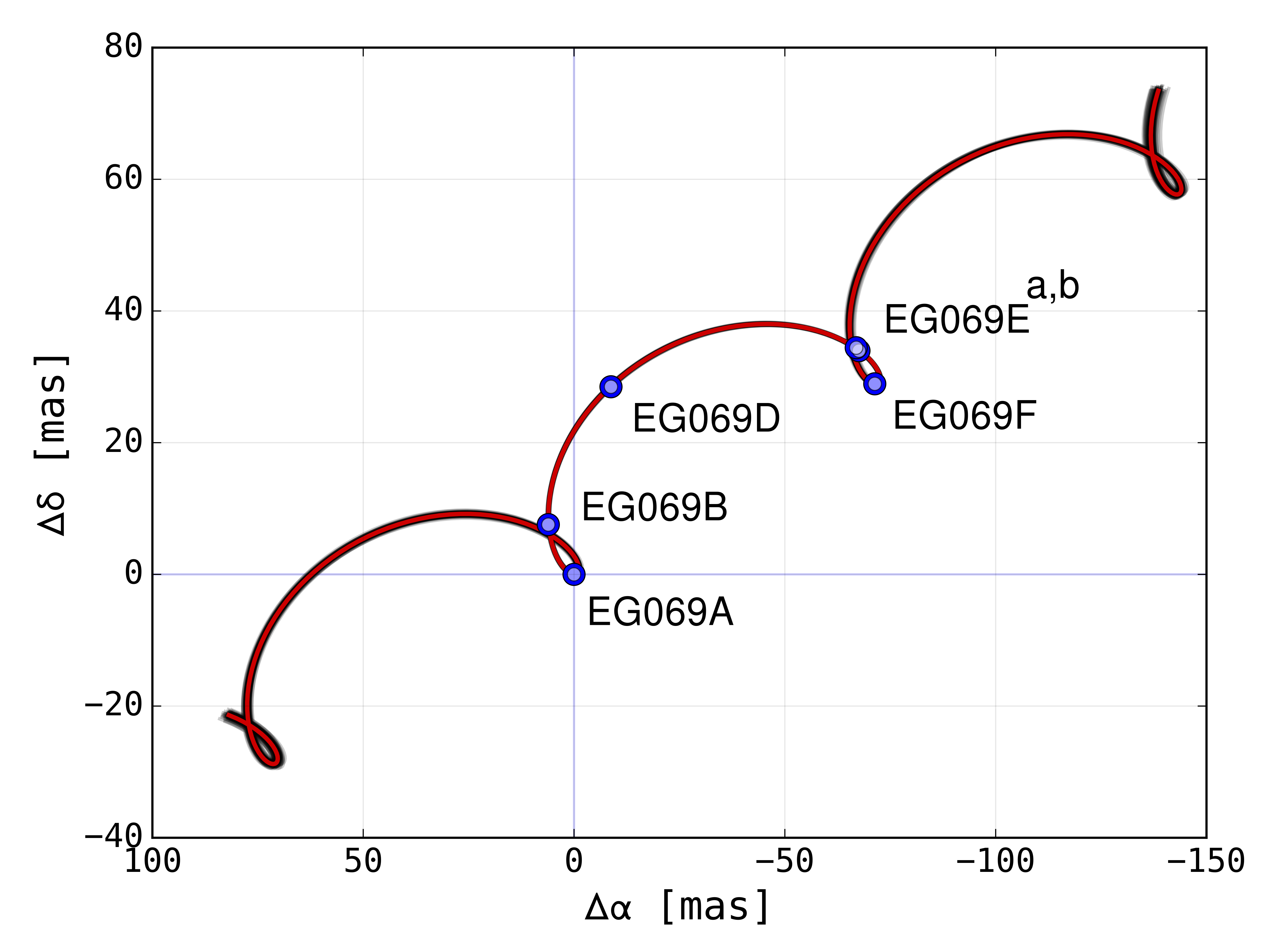}
    \includegraphics[width=0.48\textwidth]{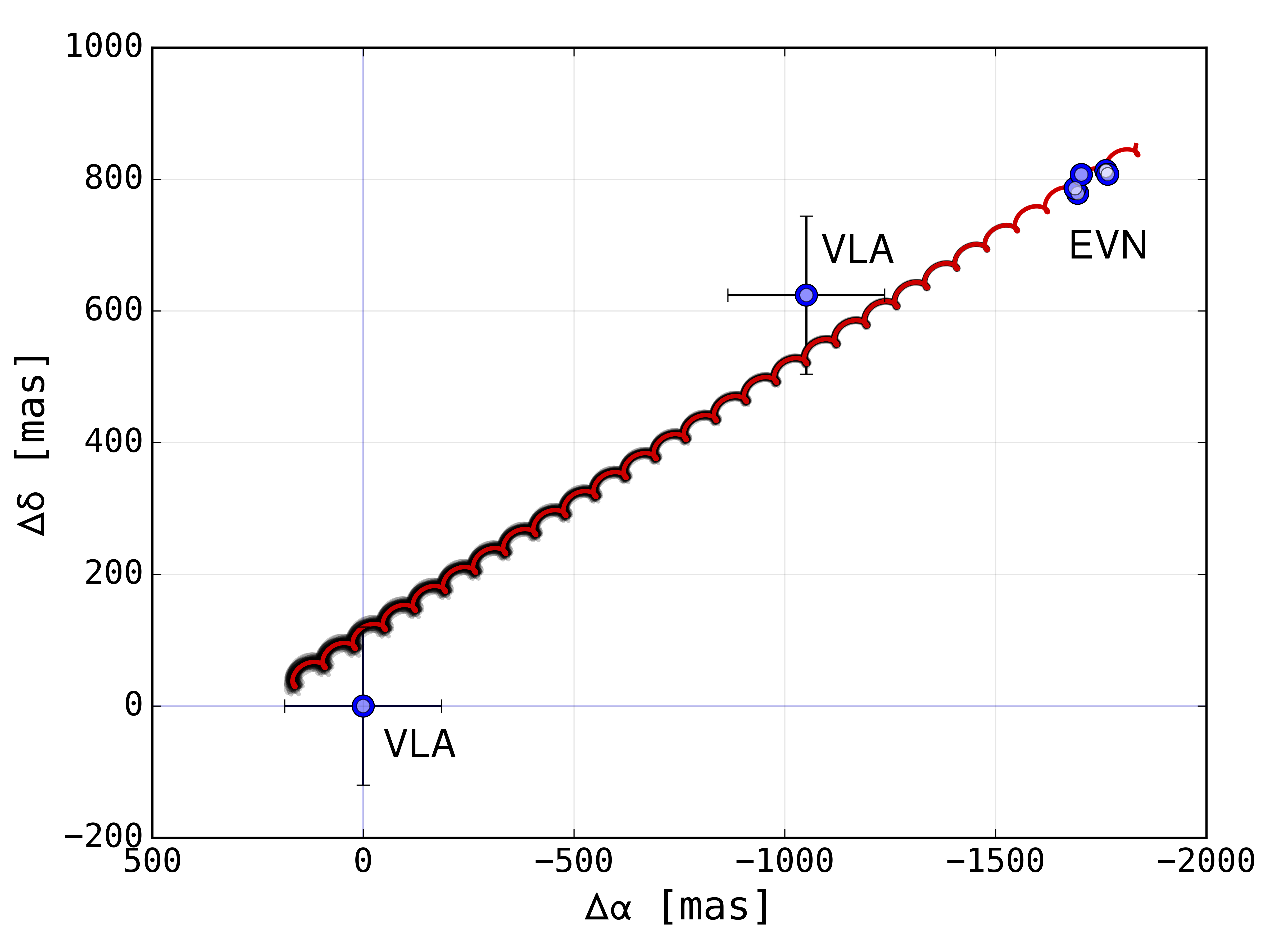}    
}    
}    
\caption{
Sky-projected paralactic motion of the target for all e-EVN observations at epochs displayed  in Tab.~\ref{tab2} ({\em top panel}),  and for all radio data ({\em bottom panel}).  Red curves are for  nominal solutions in Table~\ref{tab3}.  
Thin, grey curves are for 100 randomly selected samples from the MCMC-derived posterior, as illustrated in  Fig.~\ref{fig3}. Note that in all cases the synthetic curves are plotted for $\pm 465$~days,  prior and beyond the first and last data epoch, respectively.
}
\label{fig4}
\end{figure}  
In the best-fitting solution for all e-EVN data in Table~\ref{tab2}, the AM~Her parallax ${\rm \pi}=11.29\pm0.08$~mas,  which is equivalent to the distance of $d=88.6\pm0.6$~pc. This is  formally the most accurate and absolute determination of the AM\,Her distance, as compared to previous estimates in \citep[e.g.][]{1995A&A...303..127G,2003AJ....126.3017T}.  This new estimate roughly agrees  with the most recent determination based on the optical observations \citep[$d = 79_{-6}^{+8}$~pc;][]{2003AJ....126.3017T}, yet our uncertainty is one order of magnitude smaller.

The top panel of Fig.~\ref{fig4} illustrates  the synthetic paralactic motion of the target  over-plotted with  the original measurements as blue (grey-white) filled circles. In this scale  the error-bars are smaller than the circle diameter.  To best fitting model (red/dark grey curve) is over-plotted on 100 randomly sampled models from the  posterior data (Fig.~\ref{fig3}) for the time interval $\pm 465$~days w.r.t the first and last data epochs, respectively.

\begin{figure*}
\centerline{
\hbox{
\includegraphics[width=0.45\textwidth]{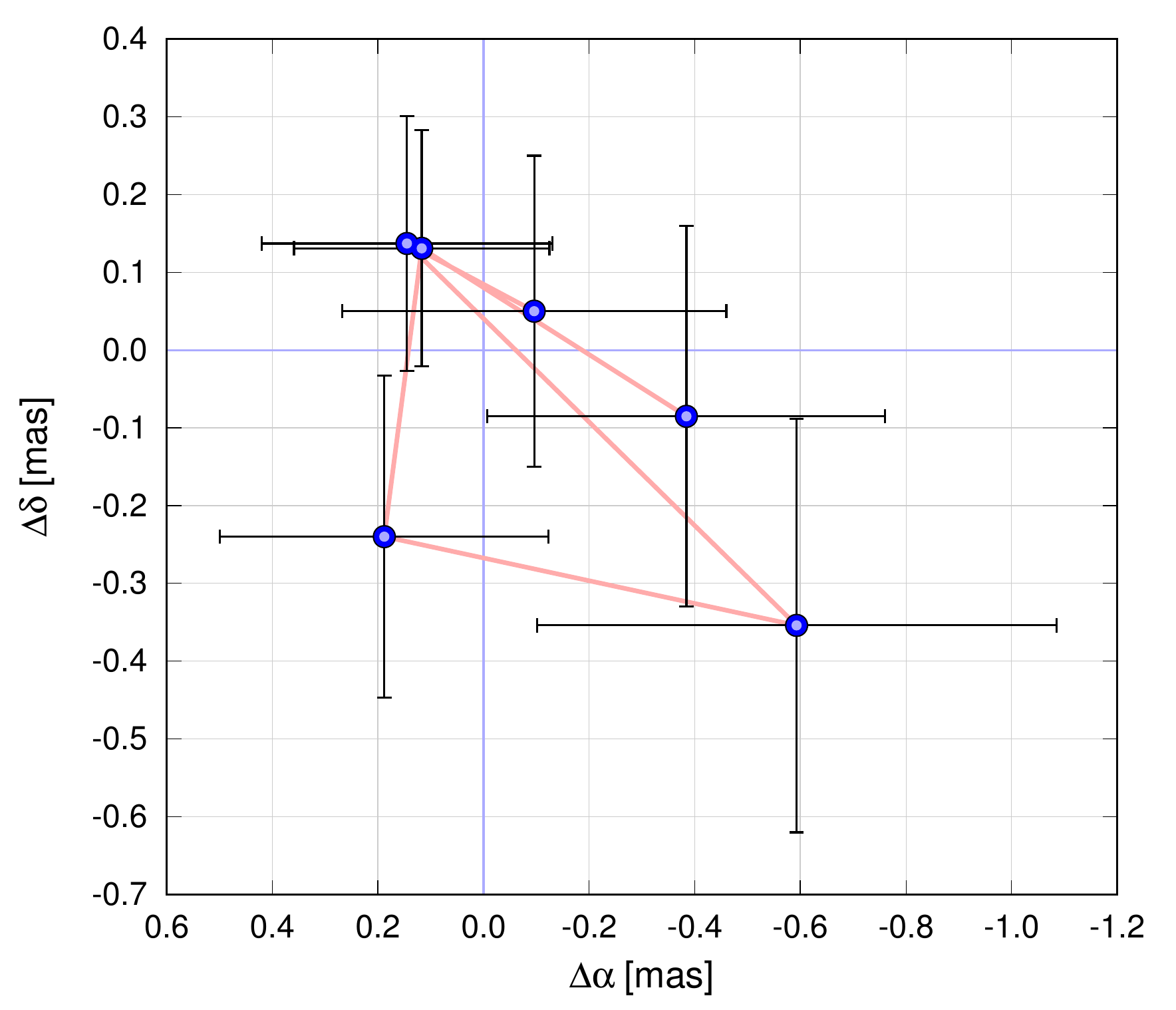}
\includegraphics[width=0.45\textwidth]{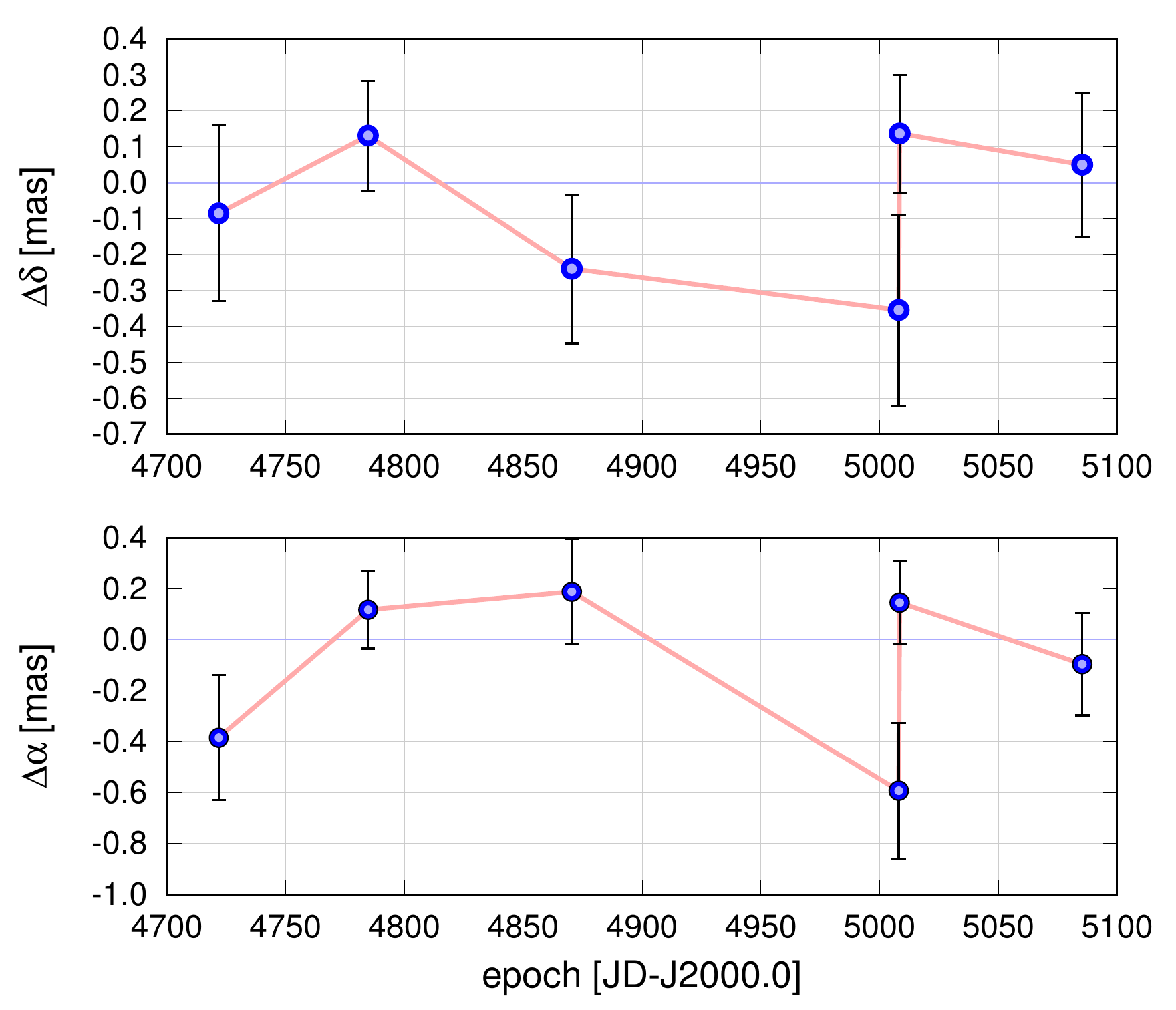}
}
}
\caption{
{\em Right}: Residuals to the best-fitting model in the $(\Delta\alpha,\Delta\delta)$-plane for all 
e-EVN measurements (see the model parameters in the left-hand column in Tab.~\ref{tab3}).
}
\label{fig6}
\end{figure*}

We also compared the inferred model positions with AM\,Her radio maps from our campaign,  and in all cases they agree well.  The residuals to the final astrometric  model based on the EVN observations in Tab.~\ref{tab2} are illustrated in Fig.~\ref{fig6}.  
Curiously, there is some systematic trend of the residuals apparent, however a sparse sampling makes it hardly possible to interpret the pattern of the residuals. Given the small error floor parameter, it has unlikely systematic meaning.

\begin{table*}
\begin{center}
\begin{tabular}{l c c c c}
\hline
%\hline
parameter   & e-EVN (6 epochs)  & e-EVN (6 epochs w.r.t. GAIA DR1) & e-EVN (4 epochs) & e-EVN+VLA (all data)    \\
\hline
\noalign{\smallskip}
\smallskip 
$\alpha_0$ & 
${18^{\idm{hr}}16^{\idm{m}}13.19074^{\idm{s}}}^{+0.00001}_{-0.00001}$ &
${18^{\idm{hr}}16^{\idm{m}}13.18336^{\idm{s}}}^{+0.00004}_{-0.00004}$ &
${18^{\idm{hr}}16^{\idm{m}}13.19074^{\idm{s}}}^{+0.00001}_{-0.00001}$ &
${18^{\idm{hr}}16^{\idm{m}}13.19074^{\idm{s}}}^{+0.00001}_{-0.00001}$ 
\smallskip\\
$\delta_0$ & 
${49^{\circ}52'5''.13685}^{+0.00006}_{-0.00006}$ &
${49^{\circ}52'5''.18152}^{+0.00028}_{-0.00028}$ &
${49^{\circ}52'5''.13684}^{+0.00007}_{-0.00007}$ &
${49^{\circ}52'5''.13685}^{+0.00007}_{-0.00007}$ 
\smallskip\\
$\mu_{\alpha}^{*}$ [mas yr$^{-1}$]&  
{-46.02}${}^{+0.22}_{-0.22}$ &
{-46.02}${}^{+0.22}_{-0.22}$ &
{-46.00}${}^{+0.19}_{-0.19}$ &
{-46.01}${}^{+0.23}_{-0.24}$ 
\smallskip\\
$\mu_{\delta}$ [mas yr$^{-1}$]&    
{28.83}${}^{+0.18}_{-0.18}$ &
{28.83}${}^{+0.18}_{-0.18}$ &
{28.83}${}^{+0.16}_{-0.16}$ &
{28.83}${}^{+0.19}_{-0.19}$ 
\smallskip\\
parallax $\pi$ [mas]   &  
{11.29}${}^{+0.08}_{-0.08}$ &
{11.29}${}^{+0.08}_{-0.08}$ &
{11.27}${}^{+0.08}_{-0.08}$ &
{11.29}${}^{+0.09}_{-0.09}$ 
\\
\noalign{\smallskip}
\hline
\end{tabular}
\end{center}  
\caption[]{
Parameters of the best-fitting solution for three data-sets including all  e-EVN epochs in Tab.~\ref{tab2} 
({\em first and second column}) at the middle-arc epoch $t_{\rm 0}=$\,JD~2456457.5 and the GAIA DR1 epoch 
JD~2457023.5, respectively; a minimal  set of four observations making it possible to determine  the parallax 
({\em the third column}) for the epoch $t_0$,  and the set or radio-interferometric data including archival 
VLA measurements from NRAO database,  ({\em right column}), also for the epoch $t_0$ , respectively. We note that position 
uncertainties for GAIA DR1 epoch (the second column) are relatively large due to parameter correlations since the initial 
epoch is outside the measurements time window. 
}
\label{tab3}
\end{table*} 

Given that the EVN observations are very expensive in terms of the human power and telescope time,  we did also 
a simple experiment for estimating a minimal number of observations required to reliably  determine the parallax 
of relatively distant,  AM Her-like targets.  We assume that observations with  sub-mas level uncertainties are 
scheduled to cover the whole  year time-window. We found that three  observations are not sufficient to determine 
the parallax. With six $(\alpha,\delta)$ datums,  the astrometric model in  Eq.~\ref{eq:m1} is closed, but we 
could not find any reliable solution using  the MCMC optimization procedure. With four observations,  the results are very  similar to the parameters obtained for the whole EVN  data set, both in terms of the numerical values (Table~\ref{tab3})  and the posterior distribution.

\begin{figure}
\centering
\includegraphics[width=0.48\textwidth]{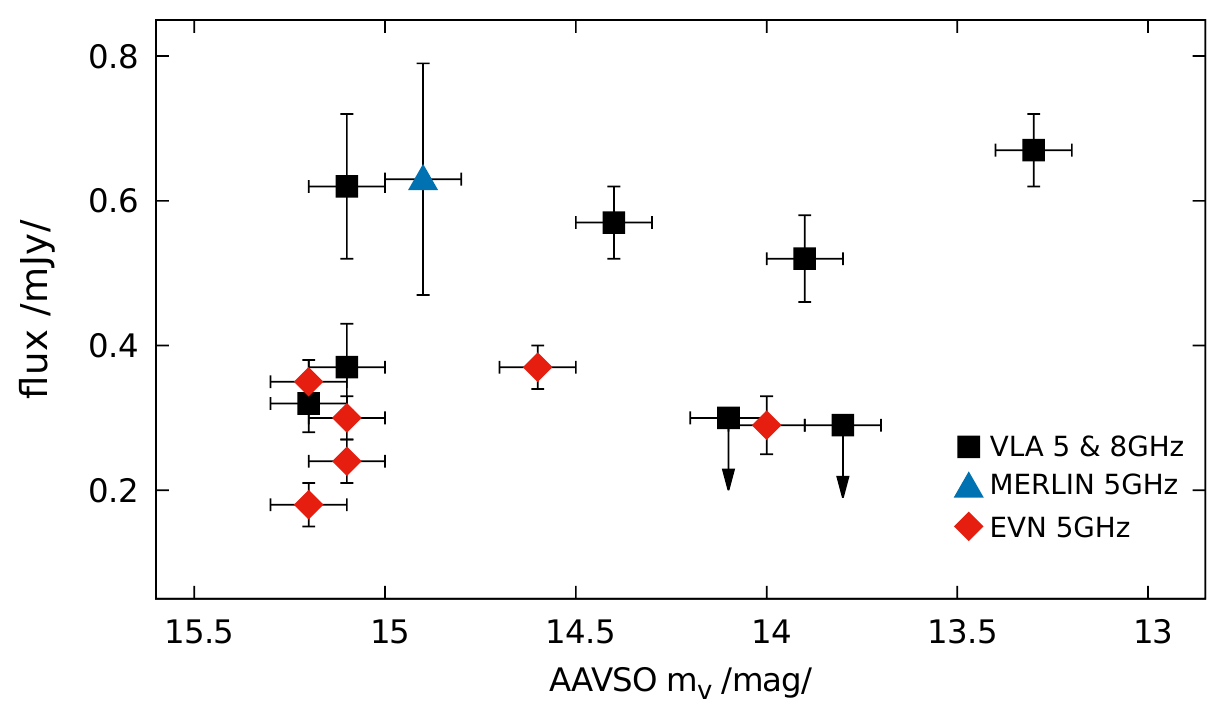}
\caption{A comparison of the measured AM\,Her radio fluxes from VLA \citep{1982ApJ...255L.107C,1983ApJ...273..249D,
1985ASSL..116..225B,2007ApJ...660..662M}, MERLIN \citep{1994MNRAS.269..779P} and EVN (this paper)
with optical observations from AAVSO data. For AAVSO data we assumed error of 0$^m$.1 for 3-day 
average around the epoch of the e-EVN observations.}
\label{fig10}
\end{figure}

Finally, we extended our e-EVN measurements with two archival VLA observations from the NRAO  
database\footnote{\url{science.nrao.edu/facilities/vlba/data-archive}}. The VLA data, back to 1988 
and 2003,  could improve the determination of the proper motion. For the same reason, we could use  
optical measurements in \cite{2003AJ....126.3017T}, however, they are not available in source form  and 
the reference paper reports large uncertainties  $\sim1''$. We also found an infra-red position  from 
the 2MASS catalogue \citep{2006AJ....131.1163S}, which has similarly substantial uncertainty 
of $\sim80$\,mas.

We added the VLA measurements aiming to improve the mean motion parameters, since the EVN data cover only one year. 
Unfortunately, the uncertainties are much larger for this set than for the e-EVN data,  and the VLA measurements 
stand out from the model (the bottom panel in Fig.~\ref{fig4}). Yet the spread of models  randomly selected from  the posterior, is quite limited for almost three decades interval. We found that the posterior distribution looks 
like the same as for the EVN data. We did not find any improvement of the mean motion parameters too, see Table~\ref{tab3}.

\section{Properties of the observed radio emission}

\begin{figure*}
\centering
\includegraphics[width=7.8cm]{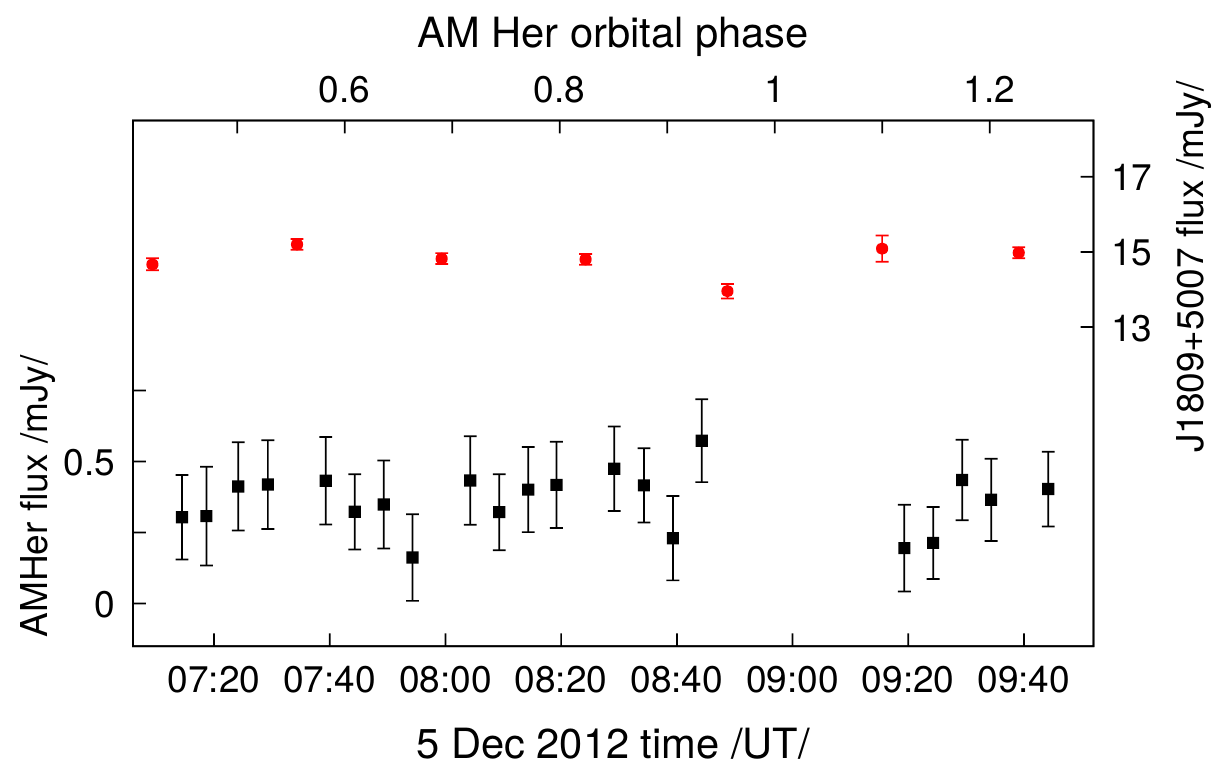} 
\includegraphics[width=7.8cm]{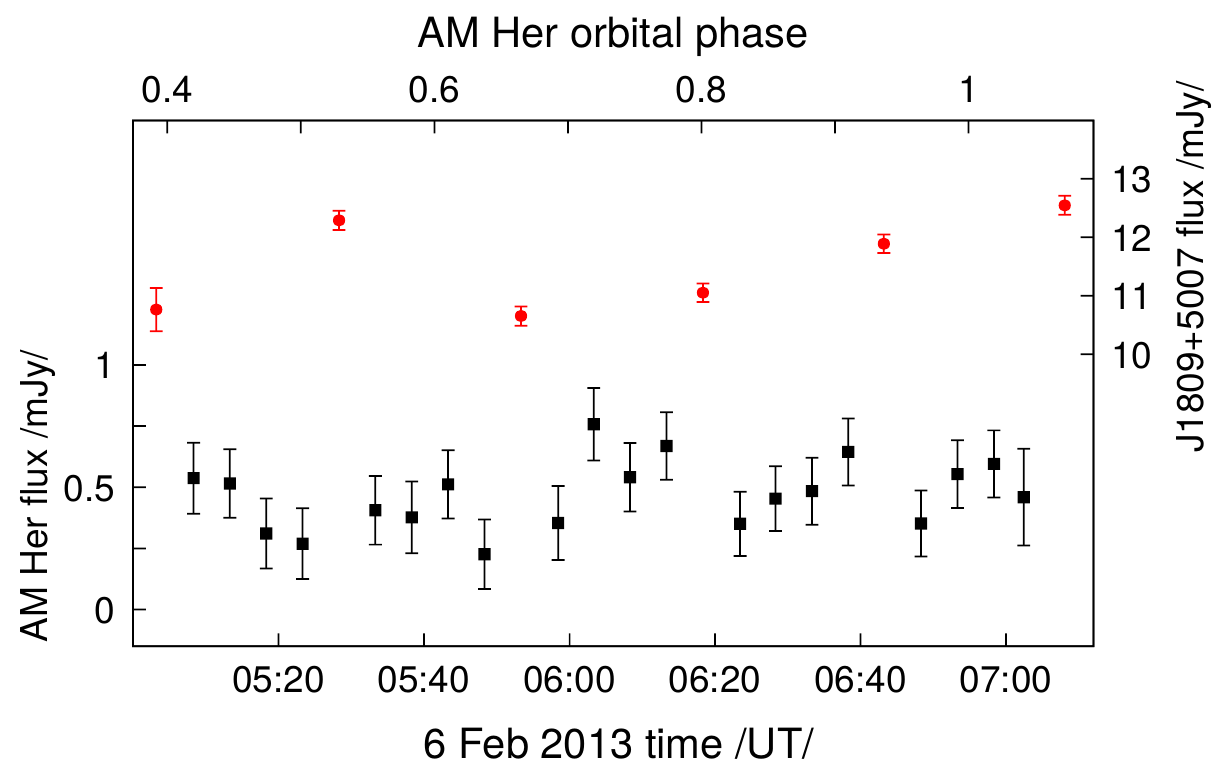}
\includegraphics[width=7.8cm]{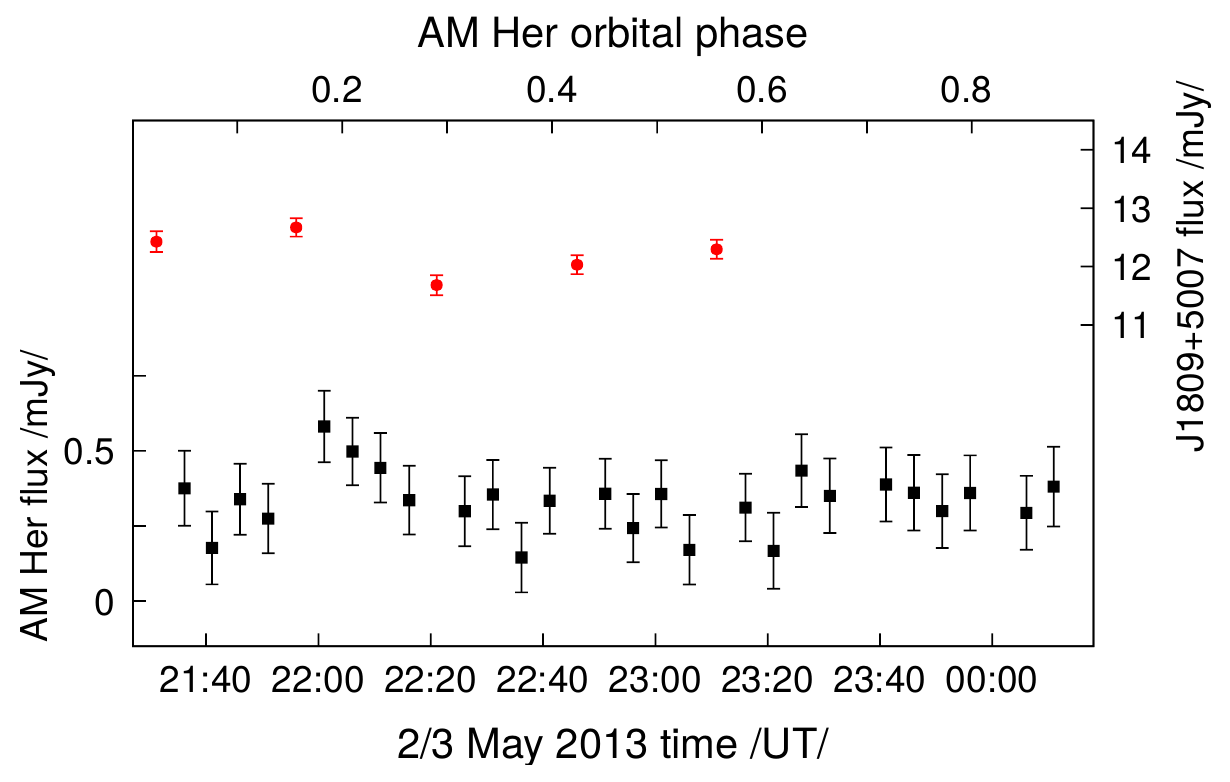} 
\includegraphics[width=7.8cm]{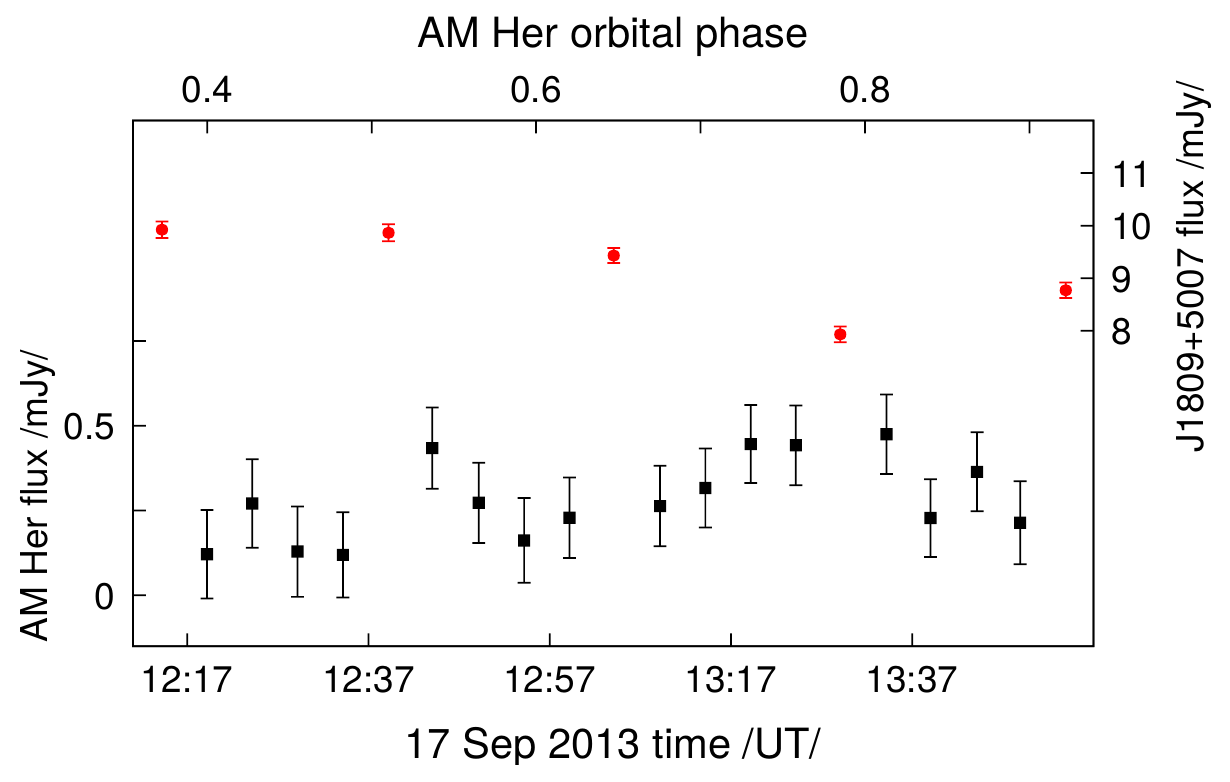} 
\includegraphics[width=7.8cm]{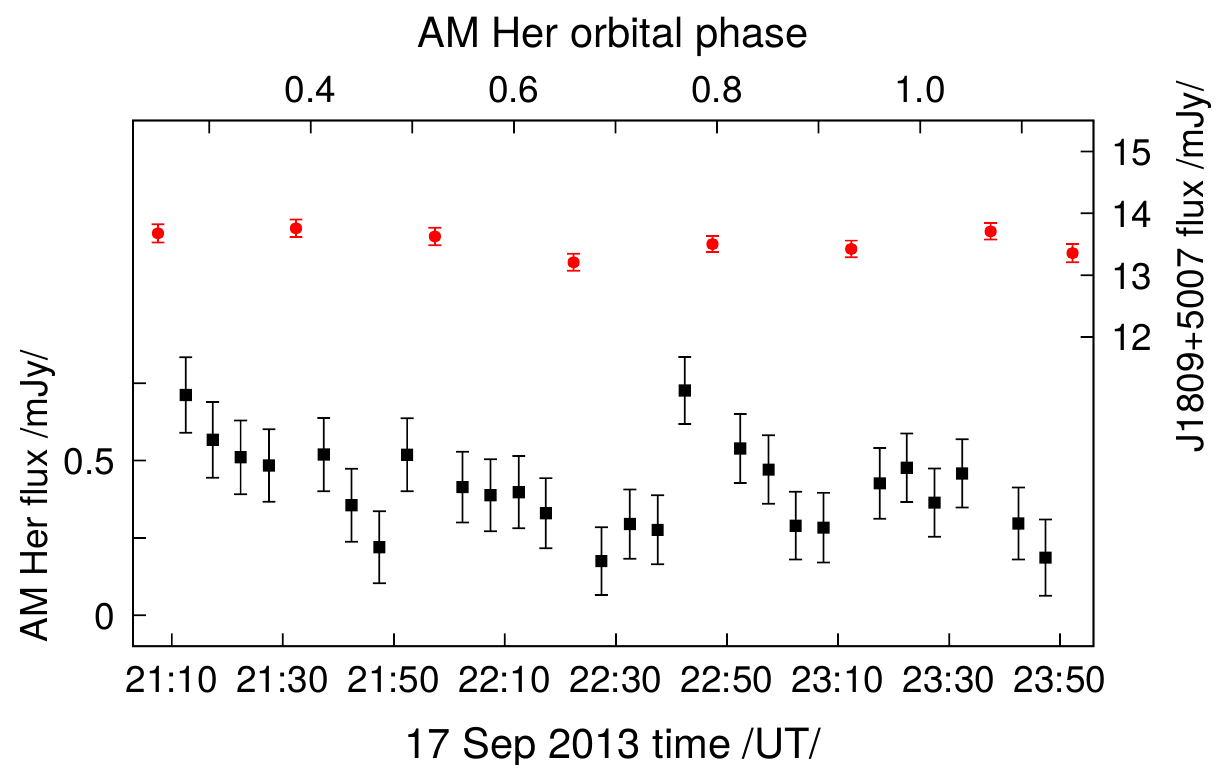} 
\includegraphics[width=7.8cm]{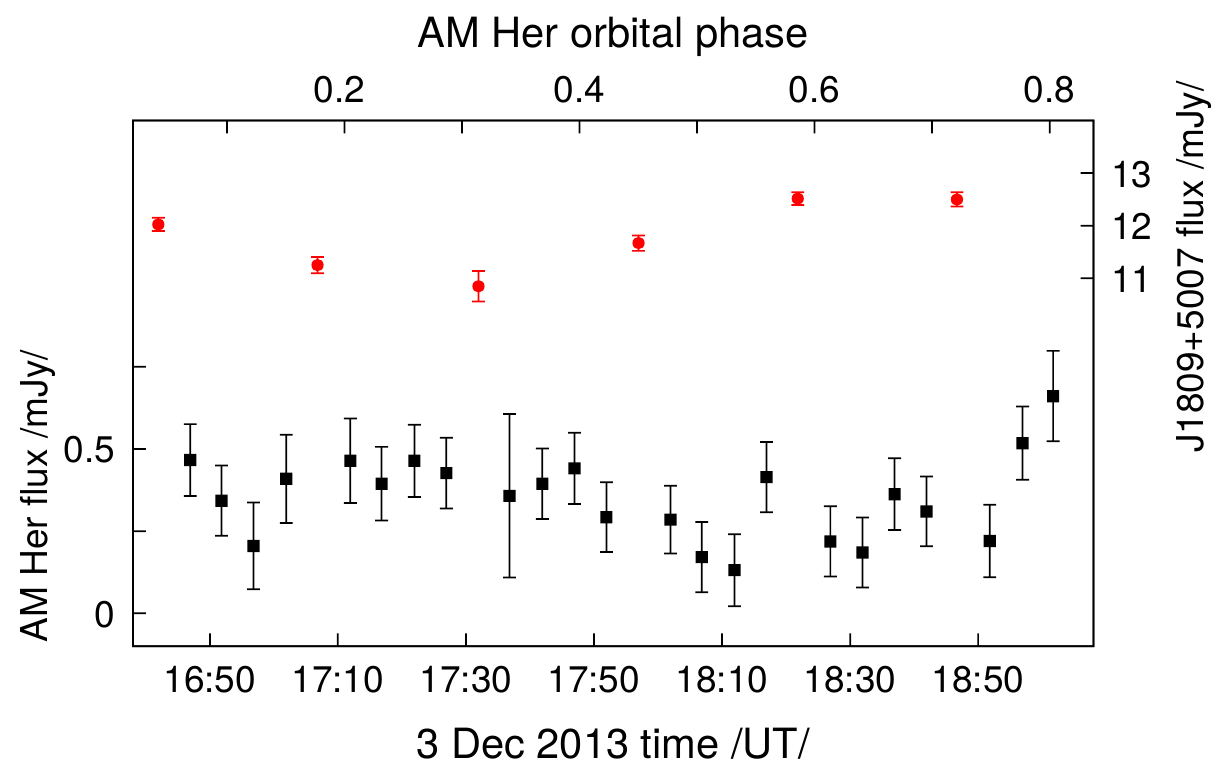} 
\caption{Variability of the radio emission obtained from our interferometric  observations. Each data 
point represents single scan during the phase-referencing observations (\textcolor{red}{$\bullet$}
represents J1809+5007 and $\mathbin{\vcenter{\hbox{\rule{0.8ex}{0.8ex}}}}$ represents AM\,Her, respectively) . The error bars are of length $\pm1\sigma$.}   
\label{fig8}
\end{figure*}

\begin{figure}
\centering
\includegraphics[width=7.8cm]{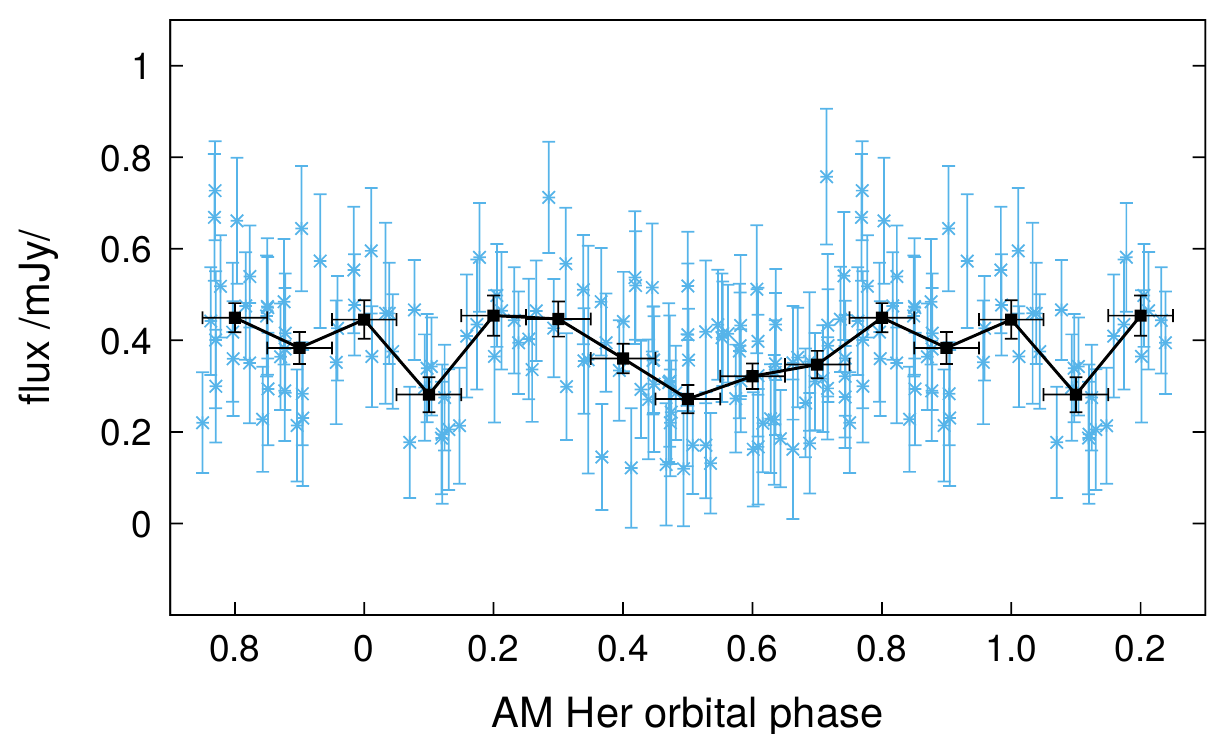} 
\caption{AM\,Her quiescent radio flux at 5\,GHz phased with the orbital motion of the system.  Cyan points
represent measurements based on individual scans and black points binned values, respectively. }   
\label{fig9}
\end{figure}

The radio emission traces particle acceleration and hence is a very useful probe of physical conditions 
in various astrophysical systems. In order to study the physical characteristics of the quiescent emission,  
we estimated the brightness temperature $T_{B}$, which in the Rayleigh-Jeans regime  can be approximated by 
\begin{equation}
\label{t_br}
T_{b}[{\rm K}]\simeq9.87\times10^{10} \frac{D^{2}_{100}\,S}{r^{2}_{11}\nu^{2}}\,,
\end{equation}
\citep{1994MNRAS.269..779P}, where $D_{100}$ is the distance in units of 100~pc,  $S$ is the radio flux density 
in mJy, $r_{11}$ is the radius of the emitting region in units  of $10^{11}$\,cm and $\nu$ is the frequency 
of the observations in units of GHz.  We calculated the maximum size of the emitting region $r_{11}$, using
estimated minimum resolvable size $\theta_{m}$ of an interferometer for Gaussian brightness distribution  
in a naturally weighted image
\citep[e.g.][]{2005AJ....130.2473K}:
\begin{equation}
\label{w2}
\theta_{m}=\sqrt{\theta_{\rm maj}\times\theta_{\rm min}}\sqrt{\frac{4\,\mathrm{ln}\,2}
           {\rm \pi}\mathrm{ln}\left(\frac{\mathrm{SNR}}{\mathrm{SNR}-1}\right)}\,,
\end{equation}
where SNR represents the signal-to-noise ratio, $\theta_{\rm min}$ and $\theta_{\rm maj}$ 
represents the major and minor axes of the restoring beam, respectively. AM\,Her 
appears unresolved on all our maps with SNR detections in the range $\sim$\,7\,--\,20. 
This implies that the minimum resolvable size is 
equal to the upper size of the emission region, and hence the lower limit of $T_{b}$ could
be estimated. Our observations give $r_{11}$=21.7\,--\,36.4 (0.15\,--\,0.25\,au), 
what translates to $T_{b}\gtrsim$\,0.4\,--2.4\,$\times10^6$\,K.  A VLBI detection with a brightness temperature 
above $10^6$\,K is usually interpreted  as a signature of non-thermal radiation, however value $\sim10^6$\,K 
do not exclude the thermal emission. Therefore, this particular estimation is ambiguous. On the other hand, 
the thermal radiation may come only from relatively large emission zone $\gtrsim30\,r_{\rm orb}$ ($T_{b}\lesssim10^6$\,K), 
where $r_{\rm orb}\simeq0.005$~au is the radius of the AM\,Her orbit. This radius was calculated under assumptions 
that the mass of the primary component is $M_{\rm WD}\simeq0.7\,$M$_{\odot}$, the secondary star mass is 
$M_{\rm RD}\simeq0.3\,$M$_{\odot}$, and the orbital period is $P_{\rm orb}=3.1$\,hr. The values of the parameters 
listed above agree with the literature data  \citep[e.g.][]{2006ApJ...639.1039G}. Such a large emission region 
could be considered in the case of the thermal bremsstrahlung in a stellar wind. However, using the wind  analysis 
of \citet{1975MNRAS.170...41W}, we estimate that the observed flux values could be achieved only for unrealistically 
high mass loss rates, i.e. for a slow wind $\varv_{\rm w}=400$~km/s, required mass loss rate is 
$\dot{M}\sim 3 \times\,10^{-8}\,$M$_{\odot}$, and for a fast wind $\varv_w=1500$~km/s, required mass loss rate 
is $\dot{M}\sim1\times\,10^{-7}\,$M$_{\odot}$. Therefore, it is unlikely that observed AM\,Her radio emission 
has thermal origin, what  is in agreement with previous conclusions \citep[e.g.][]{1983ApJ...273..249D,2007ApJ...660..662M}. 

Curiously, the radio fluxes measured during our campaign ($180-370\,\mu$Jy) appear
to be lower than those reported in the literature (so far). Moreover, our all observational
epochs took place during the decreasing and the low state of the AM\,Her optical activity.
This observational result suggests that the AM\,Her radio luminosity may be correlated with the mass transfer  
rate, which reflects in high and low states of the optical and the X-ray activity \citep[e.g.][]{2002A&A...396..213D}.  
In order to study a possible correlation between the optical and the radio activity of AM\,Her, we collected  
all flux measurements at 5 \& 8\,GHz available in the literature, and compared them with  the optical  observations 
from AAVSO archive (Fig.\ref{fig10}). We selected these particular radio bands, because the AM\,Her radio 
spectrum appears flat at these frequencies \citep{1982ApJ...255L.107C}. The high value for VLA measurement during 
the low optical state \citep[15 min integration $f_{8.4\,\mathrm{GHz}}\simeq0.63$\,mJy, $m_V\simeq15.1$, ][]{2007ApJ...660..662M} 
is most likely due to radio flare, as during the next 15 min integration, the detected flux was 
$f_{8.4\,\mathrm{GHz}}\simeq0.37$\,mJy. To test a significance of a possible correlation we
calculated the Pearson correlation coefficient for available data. We removed the probable flare
presented in \citet{2007ApJ...660..662M} from our sample and took only detections into account. 
The derived correlation coefficient $\sigma$\,=\,0.62 represents moderate correlation, where \emph{p}-value\,=\,0.03 
is a strong indication that the relationship is real.  When the non-detections from \citet{1985ASSL..116..225B} 
are added with the upper limits, we obtained $\sigma$\,=\,0.45 and \emph{p}-value\,=\,0.11. 
The \emph{p}-value indicates that we cannot reject the hypothesis that both discussed AM\,Her physical 
properties are unrelated. However, this is a small sample statistic and
more data is required to make any decisive statements on this issue.

The noted difference between the archival data and the new EVN measurements of the quiescent 
emission can be explained in two ways. First, we consider the extended, diffused component of the AM\,Her 
radio emission, that could be resolved on VLBI scales. However, it is difficult to point out its origin. 
This may be a strong stellar wind from the system, but this hypothesis again needs very high mass loss rate. 
It also could be due to a slow expanding and decelerating shell, after the nova outburst 
\citep[e.g. RS\,Oph, ][]{2009MNRAS.395.1533E}, although it requires quite a recent  ($\sim$\,1--\,10 
years ago) thermonuclear runaway event in AM\,Her. It is clearly not the case, 
since from the distance about of $\sim$\,88\,pc and with a typical absolute magnitude of  
$-\,8^{\idm{m}}$ during  the maximum  \citep{1995ApJ...452..704D}, the visual optical brightness 
of  the AM\,Her nova at the maximum should be m$_{\,max}$$\sim-3^{\idm{m}}$. This would be hardly
possible to overlook nowadays. 

Alternative scenario assumes that there is a correlation between the quiescent radio 
luminosity and the activity level of AM\,Her. \citet{1983ApJ...273..249D} proposed, that  
the quiescent emission emerges through the gyrosynchrotron process caused by mildly relativistic  
(E$\sim 500$\,keV) electrons, which are located in the magnetosphere of the primary star. Such 
a correlation implies that the accretion stream provides at least partially electrons responsible 
for the gyrosynchrotron emission, even during very low rates of the accretion. 
The relationship may not be a strong one, as other effects (e.g. local magnetic activity of 
the secondary star) also could have impact on the quiescent emission. 
We prefer this scheme from two presented, because the first one generates additional problems.

Due to the quality of the e-EVN observations we were able to track evolution  of the AM\,Her radio 
flux from short ($\sim5$\,min) to long ($\sim$\,months) timescales. 
In order to check the reliability of AM\,Her flux measurements we obtained fluxes for the phase 
calibrator J1818+5017 and the secondary calibrator J1809+5007. The fluxes for AM\,Her 
and J1809+5007 are presented on Fig.\ref{fig8}. The scatter of J1809+5007 flux from individual 
scans is within $\sim$\,10\,\% around the mean value,  and in case of J1818+5017 in $\sim$\,5\,\% 
around the mean with a few rare outliers. We also compared fluxes based on the upper and the lower
half of the frequency band and these agree within the error with a few points with noticeable 
deviation. It gives us the confidence that measured AM\,Her fluxes are valid. 

We did not detect any  short-time radio outbursts, which were 
previously reported by \citet{1983ApJ...273..249D}. We observed only flux variations around 
an average value, which may reflect short-time changes of AM\,Her radio luminosity. We also checked 
if the radio emission is modulated with the orbital period. The phase-resolved light curve is showed on Fig.\ref{fig9}. 
We used AM\,Her orbital ephemerids taken from \citet{2005AJ....130.2852K} for the calculations. Two minima are visible 
in the phased radio flux, a sharp one around $\phi\sim0.1$, and a wider one with the local minimum at $\phi\sim0.6$. 
We checked reliability of the phased radio light curve and repeated measurements for the
upper and the lower half of the used radio band. In all cases the both mentioned minima were visible and located at 
the same orbital phase. This is a strong indication that the noted dependency between the orbital phase
and the quiescent radio luminosity is real, however new more sensitive observations
are needed to support this finding. We also investigated if presented pattern in the phase-resolved light curve
could arise from a sample of random data. We calculated reduced $\chi^2$ for the binned data relative
to the mean based on all individual measurements, which gives $\chi^2\simeq2.6$.
Next we derived $\chi^2$ for 10000 iterations for binned data with randomized time for each single measurement. 
We found that a random set of data may give $\chi^2 >2.5$, with the probability that is less than 1\%. We conclude 
it is unlikely that the observed light curve could be a result of accidental measurements.

This observational result contradicts the proposed by 
\citet{1983ApJ...273..249D} explanation of AM\,Her quiescent radio emission. The model assumes that 
the emission region is comparable or larger than the physical size  of the binary. The new e-EVN 
data suggest that there is a correlation  between the observed radio flux and the AM\,Her orbital 
phase. Moreover, the radio light curve is similar  to observed in V471\,Tau \citep{1999ApJ...519..850N}, 
a pre-CV eclipsing binary with orbital period 12.51\,hr\,. \citet{1999ApJ...519..850N} proposed that 
the emission mechanism similar to RS\,CVn  binary systems could explain observed radio  properties of 
V471\,Tau, where the gyrosynchrotron  emission originates from $\sim400$\,keV electrons near 
the surface of the secondary component. This  model assumes that electrons are accelerated to 
mildly relativistic energies, in the region where the magnetic fields of both stars are reconnecting. 
The accelerated electrons trapped in the K dwarf's magnetosphere
are responsible for the radio emission. This interaction of fields is caused by a differential 
rotation of both components in V471\,Tau. The radio emission arises in a wedge-like magnetic structures, 
which connects the acceleration region with the photosphere of the secondary component 
\citep[see for details][]{1999ApJ...519..850N}. It should be noted that the observed
V471\,Tau flux variations are much more prominent in comparison to AM\,Her. This could be
just a pure geometrical effect, as in V471\,Tau there is an eclipse of the radio emitting
region by the K1V secondary photosphere. As the AM\,Her orbital inclination is relatively high
\citep[$i\simeq50^{\circ}$, e.g. ][]{1996MNRAS.280..481D} the probable dependency between 
radio flux and the orbital phase likely arises only due to different orientation of the magnetic
field structures in AM\,Her and the line of sight. If this model is valid in the case of AM\,Her, 
one important condition should be met, the secondary red dwarf should have strong ($\sim\,$kG) 
large scale magnetic field. 

Recently, \citet{2015ApJ...815...64W} showed that the well-studied  M9 type dwarf TVLM 513–46546 hosts 
a stable, dipole magnetic field of about 3~kG at the surface.  Therefore, it is plausible to assume 
that other low-mass red dwarfs are also able to create  such strong magnetic fields. The flaring and 
the spectroscopic activity of  the red dwarf observed during the low states of AM\,Her 
\citep{2005AJ....130.2852K,2006AJ....131.2673K}  supports the idea of the strong magnetic fields on 
the red dwarf surface, because the flaring is a sign post of the magnetic and star-spot activity. 
\citet{2006AJ....131.2673K} also concluded that the observed spectroscopic variations in  H$\alpha$ 
profiles are consistent with motions in large loop magnetic coronal structures on  the secondary star. 

However, the synchronous rotation of AM\,Her components causes problems for V471\,Tau model and 
the process of electron acceleration should be different in AM\,Her. We postulate that the acceleration 
may originate from the interaction between red dwarf  local magnetic fields frozen into transferred 
plasma and the white dwarf magnetosphere. The magnetic reconnection takes place near the L1 point, 
where the plasma accumulates during  the accretion.  As the local magnetic activity is very variable 
in the case of active red dwarfs, it may naturally explain observed variations in the quiescent radio 
flux in short and long timescales.  This is also in agreement with the observed probable correlation between 
the radio luminosity and the high and low states of AM\,Her activity, when during the increased 
mass-transfer rate the electron  reservoir is simply much larger  \citep[$\sim 3\times 10^{-11}\,{\rm M}_{\odot}$\,{\rm yr}$^{-1}$ during  high state and at least one order  of magnitude lower during low state, ][]{1998A&A...333L..31D}. Observed variations in  the quiescent radio flux on timescales of minutes/hours may be also interpret 
as changes in the mass  transfer rate. Such rapid changes in the accretion are observed in the optical 
and X-ray domain  \citep[e.g.][]{1998A&A...333L..31D,2000A&A...354.1003B}

\citet{2007ApJ...660..662M} discovered a second persisted radio polar AR Ursae Majoris (AR\,UMa), which 
has different physical properties than AM\,Her. AR\,UMa is a binary system with the orbital period 
1.93\,hr \citep{1994ApJ...426..288R}. The primary white dwarf in this system has the magnetic field 
strength of about 230\,MG \citep{1996ApJ...473..483S}, and its mass is in the range $0.91 - 1.24\,$M$_{\odot}$  
\citep{2016ApJ...828...39B}. \citet{2005ApJ...632L.123H} using infrared spectroscopy, estimated the spectral 
type of the AR\,UMa secondary red dwarf (M5.5\,V). \citet{2007ApJ...660..662M} also noted that AR\,UMa 
phased radio light curve at 8.4\,GHz suggests a minimum near the orbital phase $\phi\sim0$. If we assume 
that the source of the quiescent radio emission is the same in both polars, it leads to a conclusion that 
the emission does not depend on the physical properties of the primary white dwarf.  Our findings support 
\citet{2007ApJ...660..662M} statement, where  the authors postulate that the quiescent radio emission is 
a sign-post of a magnetized secondary star and this distinguish AM\,Her and AR\,UMa from other polars, where 
no such emission was detected.  A more precise phased radio light curve of AR\,UMa of both systems in high 
and low states of activity would shed a new light on this puzzle.

\section{Conclusions}
We report our results from the recently conducted e-EVN astrometric campaign at 6\,cm of AM\,Her. These observations  
were conducted in years 2012\,--\,2013. AM\,Her was detected on all six scheduled observational epochs with the quiescent  
radio flux in the range  0.18\,--\,0.37\,mJy. We calculated a new AM\,Her astrometric model, and we determined an improved 
annual, absolute parallax of $\pi=11.29\pm0.08$\,mas with the uncertainty one order of magnitude less than in the literature. 
It places the AM Her almost 10~pc $\equiv 10\%$  farther than predicts the most recent estimate by \citet{2003AJ....126.3017T}. 
The sub-mas accuracy of the derived astrometric positions may be similar to the outcome expected from the GAIA mission, and
our results could be used as an independent technique for the GAIA measurements. We demonstrated that the e-EVN makes 
it possible to measure the AM Her parallax with only four epochs during one year interval, still providing very low uncertainty.

We found observational evidence that the AM\,Her radio flux is likely modulated  with the orbital phase and its behavior 
resembles the radio  light curve noticed in V471\,Tau. This behavior  could be explained when the origin of AM\,Her radio 
emission is similar to proposed for V471\,Tau and generally for RS\,CVn. We also postulate that the quiescent radio emission 
distinguish AM\,Her and AR\,UMa from other polars,  as systems with magnetized red dwarfs.  We also proposed that the correlation 
between the quiescent radio luminosity and the mass transfer rates (high and low states of activity) could explain noted 
difference between the AM\,Her flux based on our new EVN  observations and the archival data. This may indicate that the accretion 
stream provides electrons, which are further accelerated and produce photons in the gyrosynchrotron process, but new 
sensitive radio observations of AM\,Her during high state and AR\,UMa in both activity states are needed to validate 
this hypothesis. 

\section*{Acknowledgments}
We are grateful to Polish National Science Centre for the financial support (grant no.  2011/01/D/ST9/00735).  The EVN is a joint 
facility of European, Chinese, South  African, and other radio astronomy institutes funded  by their national  research councils. 
K.G. gratefully acknowledges the Poznan Supercomputing and Networking  Centre (PCSS, Poland) for continuous support and computing 
resources through grant No. 313. This research has  made use of the SIMBAD database operated  at CDS, Strasbourg, France. 
We acknowledge with thanks the variable  star observations  from the AAVSO International Database contributed by observers 
worldwide and used in this research.

\bibliographystyle{mnras} % style aa.bs
\bibliography{ms} % your references Yourfile.bib
\bsp

\label{lastpage}
\end{document}